\documentclass[12pt]{article}
\usepackage{amsmath}
\usepackage{multirow}
\usepackage{amsfonts}
\usepackage{amssymb,graphics,psfrag}
\usepackage{array,epsfig,multirow,graphicx}
\usepackage{comment}
\usepackage{slashed}
\usepackage{hyperref}
\usepackage{tensor}
\usepackage{booktabs}
\usepackage{colortbl}
\usepackage{hhline}
\usepackage{mathtools}
\usepackage{enumitem}
\usepackage{mathrsfs}  
\usepackage{xfrac}
\usepackage{tikz}
\usepackage{amssymb}
\usepackage[numbers,sort&compress]{natbib}
\usetikzlibrary{decorations.markings}
\usetikzlibrary{shapes,arrows}
\usetikzlibrary{decorations.pathreplacing}
\usetikzlibrary{arrows,positioning}
\usepackage{verbatim}
\usepackage{makecell}
\usepackage{setspace}
\usepackage{anyfontsize}
\usepackage{ctable}
\usepackage{braket}
\usepackage{arydshln}

\def\hybrid{\topmargin -20pt    \oddsidemargin 0pt
        \headheight 0pt \headsep 0pt 
        \textwidth 6.25in      
        \textheight 9 in      
        \marginparwidth .875in
        \parskip 5pt plus 1pt
          \jot = 1.5ex
  }
\hybrid
\numberwithin{equation}{section}
\numberwithin{table}{section}

\newcommand{\beq}{\begin{equation}}
\newcommand{\eeq}{\end{equation}}
\newcommand{\be}{\begin{equation}}
\newcommand{\ee}{\end{equation}}
\newcommand{\bea}{\begin{eqnarray}}
\newcommand{\eea}{\end{eqnarray}}
\newcommand{\ben}{\begin{eqnarray*}}
\newcommand{\een}{\end{eqnarray*}}               
\newcommand{\ba}{\begin{align}}
\newcommand{\ea}{\end{align}}
\newcommand{\bt}{\begin{tabular}}
\newcommand{\et}{\end{tabular}}
\newcommand{\bc}{\begin{center}}
\newcommand{\ec}{\end{center}}
\newcommand{\ax}{\alpha}

\newcommand{\cO}{\mathcal{O}}

\newcommand{\cN}{\mathcal{N}}

\newcommand{\cI}{\mathcal{I}}

\newcommand{\nn}{\nonumber}

\newcommand{\cref}{{\bf [check ref]}}

\newcommand{\tr}{\mathrm{tr}}

\definecolor{mppgreen}{RGB}{17,102,86}
\definecolor{mppgray}{RGB}{221,222,214}

\def\e{\text{e}}

\def\tbz{{\scriptscriptstyle{(0)}}}

\def\fr{\frac}

\def\Z0{Z^\tbz}
\def\bs{\boldsymbol}
\def\t{\text}

\def\tr{\text{tr}\,}

\def\bs{\boldsymbol}

\def\lab{\label}

\def\blfootnote{\xdef\@thefnmark{}\@footnotetext}
\long\def\symbolfootnote[#1]#2{\begingroup%
\def\thefootnote{\fnsymbol{footnote}}\footnote[#1]{#2}\endgroup}

\begin{document}

\baselineskip=15pt

\begin{titlepage}
\begin{flushright}
\parbox[t]{1.8in}{\begin{flushright} ~ \\
~ \end{flushright}}
\end{flushright}

\begin{center}

\vspace*{ 1.2cm}

\begin{spacing}{2.5}
\bf{{{\fontsize{24}{1}\selectfont  Black Holes and (0,4) SCFTs from Type IIB on K3}}}
\end{spacing}

\vskip 1.2cm

\renewcommand{\thefootnote}{}
\begin{center}
 {Christopher Couzens, Huibert het Lam, Kilian Mayer and Stefan Vandoren 

 \footnotetext{c.a.couzens@uu.nl~ \ ~ h.hetlam@uu.nl~ \ ~ k.mayer@uu.nl~ \ ~  s.j.g.vandoren@uu.nl}}
\end{center}
\vskip .2cm

{
Institute for Theoretical Physics and \\
Center for Extreme Matter and Emergent Phenomena,\\
Utrecht University, Princetonplein 5, 3584 CE Utrecht, The Netherlands\vspace{0.5cm}
}

\vspace*{.2cm}

\end{center}

 \renewcommand{\thefootnote}{\arabic{footnote}}
 
\begin{center} {\bf Abstract } \end{center}

\noindent 
We study the central charges and levels of 2d $\mathcal{N}=(0,4)$ superconformal field theories that are dual to four- and five-dimensional BPS black holes in compactifications of type IIB string theory on a K3 surface. They arise from wrapping a D3-brane on a curve $C$ inside K3 and have transverse space either an ALE or ALF space. These D3-branes have an AdS$_3 \times \mathrm{S}^3/\Gamma$ near horizon geometry where $\Gamma$ is a discrete subgroup of $SU(2)$. We compute the central charges and levels of the 2d SCFTs both in the microscopic picture and from six-dimensional $\cN=(2,0)$ supergravity. These quantities determine the black hole entropy via Cardy's formula. We find agreement between the microscopic and macroscopic computations. The contributions from one-loop quantum corrections to the macroscopic result are crucial for this matching.

\end{titlepage}

\tableofcontents

\newpage




\section{Introduction}

The microscopic counting of the Bekenstein--Hawking entropy of BPS black holes in string theory has had a long and rich history since its inception in \cite{Strominger:1996sh}. The authors of \cite{Strominger:1996sh} considered five-dimensional non-spinning black holes arising from the compactification of type II string theory on K3$\times \text{S}^1$, matching the entropy of the macroscopic configuration with the microscopic one. This was later extended to five-dimensional non-spinning black holes in M-theory on a Calabi--Yau threefold \cite{Vafa:1997gr} and four-dimensional black holes in M-theory on the product of a Calabi--Yau threefold with a circle by MSW \cite{Maldacena:1997de}. 
More recently black holes in F-theory \cite{Vafa:1996xn} have been given further consideration in \cite{Haghighat:2015ega,Bena:2006qm,Grimm:2018weo} for example.


In this paper we are interested in compactifications of type IIB string theory on K3 and the black strings emerging in this theory. The goal is to match macroscopic data with a complimentary microscopic description. The literature on the topic is vast, in particular for black holes arising from the D1-D5-(p) system \cite{Strominger:1996sh,Callan:1996dv}. We shall instead concern ourselves with black strings arising from wrapping D3-branes on a curve inside K3, which, in comparison to the D1-D5-(p) system, has been far less studied. Explicitly we shall consider D3-branes in an asymptotic geometry $\mathbb{R}\times \text{S}^{1} \times M \times \text{K3}$ where the D3-branes are wrapped on $\text{S}^{1} \times C$ with $C$ a curve in K3 and probe the transverse space $M$. The equations of motion and supersymmetry preservation imply that the four-manifold $M$ is a Ricci-flat hyper-K\"ahler manifold which we take to be non-compact. We consider two families of such manifolds $M=M_\Gamma$: asymptotically locally Euclidean (ALE) and asymptotically locally flat (ALF) spaces, both of which may be defined by a choice of discrete subgroup $\Gamma \subset SU(2)$ as we review in section \ref{sec:6dstrings}. Different choices of $\Gamma$ lead to black strings with different charges, some of which have not been considered before in the literature. In particular, black strings arising from D3-branes probing $M_\Gamma$ corresponding to the D- and E-series within the ADE-classification of discrete subgroups of $SU(2)$ have not been studied previously. Microscopically the strings are dual to (0,4) superconformal field theories (SCFTs) which admit both left- and right-moving central charges $c_{L,R}$. In addition they always admit a right-moving $SU(2)_R$ current algebra and in the case of the A-series also a left-moving $U(1)_L$ current algebra, each with an associated level. From this data the entropy of the black strings follows via Cardy's formula. In this paper we compute the central charges and levels corresponding to the settings described above both from a macroscopic and microscopic viewpoint.

Macroscopically we shall work exclusively in six-dimensional $\mathcal{N}=(2,0)$ supergravity \cite{Romans:1986er,Riccioni:1997np} which is obtained by compactifying type IIB on K3 \cite{Townsend:1983xt}. The techniques for computing the various central charges and current algebra levels corresponding to the black string solutions of this theory were developed in \cite{Kraus:2005vz,Kraus:2005zm,Hansen:2006wu,Dabholkar:2010rm,Haghighat:2015ega,Grimm:2018weo}.  Concretely, we reduce the classical six-dimensional action to three dimensions in the black string background. The central charges and levels correspond to coefficients of Chern--Simons terms in the three-dimensional effective action. When $M_\Gamma$ is ALF one also has to include one-loop Chern--Simons terms that arise from integrating out massive Kaluza-Klein modes \cite{Grimm:2018weo}. These macroscopic calculations are carried out in section \ref{sec:macro}.

Alternatively, we could have chosen to look at these solutions directly in type IIB supergravity and compute the central charges there. The geometries of wrapped D3-brane solutions giving rise to AdS$_3$ near horizons were classified in \cite{Kim:2005ez}. This was further extended in \cite{Couzens:2017way, Couzens:2017nnr} to include seven-branes in addition to the D3-branes. The analysis performed in \cite{Couzens:2017way} shows that for wrapped D3-brane solutions preserving $\mathcal{N}=(0,4)$ supersymmetry the geometry is essentially unique and takes the form AdS$_{3}\times$S$^{3}/\Gamma\times$ CY$_{2}$. The central charges for certain 2d CFTs were computed in \cite{Couzens:2017way} and one could in principle perform similar computations here, however we will work exclusively in six dimensions.


Microscopically the central charges and levels follow from the construction of new 2d $\mathcal{N}=(0,4)$ SCFTs. When $M_\Gamma$ is ALE these arise from wrapping known 4d $\mathcal{N}=2$ quiver gauge theories on a Riemann surface. Wrapping 4d $\mathcal{N}=2$ quiver theories on Riemann surfaces are in themselves not new, see for example \cite{Putrov:2015jpa}. However, in the context of black hole microstate counting this is the first instance of such a 2d construction. The 4d parent quiver gauge theories are given by projections of non-abelian $\mathcal{N}=4$ supersymmetric Yang--Mills preserving $\mathcal{N}=2$ supersymmetry \cite{Douglas:1996sw,Johnson:1996py}. One then wishes to place the theory on a curve $C$ inside K3. In order to preserve $\mathcal{N}=(0,4)$ in 2d this requires a particular topological twist to be performed. We then compute the central charges and levels and find a perfect matching to the macroscopic computations. When $M_\Gamma$ is ALF and $\Gamma=\mathbb{Z}_m$ corresponding to the A-series, we can compute the central charges and levels by considering a dual M-theory setting. In this case $M_\Gamma$ is the Taub-NUT space and we can T-dualize along the NUT-circle to obtain a type IIA setting. Lifting this to M-theory results in M-theory on K3$\times T^2$ with an M5-brane that wraps $C\times T^2$ and $m$ M5-branes wrapping K3. These M5-branes combine into a single M5-brane wrapping $C\times T^2 +m$K3 when the corresponding class is very ample. From this duality frame it is possible to determine the central charges \cite{Maldacena:1997de,Lambert:2007is,Dabholkar:2010rm}. For the D-series such a clean description is not possible, we shall comment on this case further later.


In section \ref{sec:6dstrings} we begin by outlining the various setups considered in this paper, in the process reviewing ALE and ALF spaces. Section \ref{sec:macro} is devoted to the computation of the macroscopic central charges and levels of the 6d strings considered here. Beginning with a discussion on six-dimensional $\mathcal{N}=(2,0)$ supergravity and its relevant black string solutions, section \ref{sec:macro} proceeds with the calculation of the classical and quantum contributions to the central charges and levels before a summary section collating the macroscopic results. The complimentary microscopic calculation of the central charges and levels, split between ALE and ALF spaces, is performed in section \ref{sec:microscopics}. We find perfect agreement with the macroscopic results of section \ref{sec:macro}. We summarize our results in section \ref{sec:conclusions}. We relegate some technical material to three appendices.

\section{Setup: black holes, strings and ALE/ALF spaces}\label{sec:6dstrings}

In this section we shall give an overview of the various setups that we consider in this paper. All the cases considered here have the same underlying theory arising from compactifying type IIB string theory on a K3 surface. In the low energy limit this leads to an effective six-dimensional supergravity theory with chiral $\mathcal{N}=(2,0)$ supersymmetry, see section \ref{sec:6dsugra} for further details. 

We shall consider D3-brane states giving rise to strings in the six-dimensional effective theory. As such the R-R and NS-NS fields that couple to D1, F1, D5 and NS5-branes will be switched off as we explain in section \ref{sec:6dsugra}. To obtain strings in six dimensions we wrap the D3-branes on a genus $g$ Riemann surface $C \subset \text{K3}$.  When the non-compact transverse space to the string in 6d is taken to be $\mathbb{R}^{4}$ the effective two-dimensional worldvolume theory living on the D3-branes is known to flow in the IR to a 2d $\mathcal{N}=(4,4)$ SCFT \cite{Bershadsky:1995vm}. The $SO(4)_{\perp}=SU(2)_L \times SU(2)_R$ rotation group of the transverse $\mathbb{R}^{4}$ realizes holographically the left- and right-moving $SU(2)_{L,R}$ R-current algebras with associated levels $k_{L,R}$. One should contrast this with the setups which form the basis of this paper and to which we now turn.

We now replace the transverse $\mathbb{R}^{4}$ with a different Ricci-flat hyper-K\"ahler non-compact four-manifold.  Such spaces have been classified into four categories depending on their asymptotic volume growth: ALE, ALF, ALG, ALH. At infinity the metrics approach a quotient of the flat metric on $\mathbb{R}^{4-k}\times T^{k}$ where the fibration at infinity is trivial except in the $k=1$ case where it may be fibered. The metric is ALE when $k=0$, ALF when $k=1$, ALG when $k=2$ and finally ALH when $k=3$. We shall only consider two of these classes in the following work: \emph{asymptotically locally Euclidean} (ALE) and \emph{asymptotically locally flat} (ALF) spaces\footnote{Note the the `G' and `H' do not have a meaning like in the `E' and `F' cases but are named as such by induction.}. For the benefit of the reader not familiar with these manifolds, or in need of a refresher, we summarize the salient points here. 

A manifold is said to admit an ALE metric if it is diffeomorphic to $\mathbb{R}^{+}\times \text{S}^3/\Gamma$, with $\Gamma$ a freely acting discrete subgroup of $SU(2)$, and has a metric that asymptotically approaches a quotient of the Euclidean flat space metric. The possible subgroups of $SU(2)$ admit an ADE-classification and are summarized in table \ref{ADE}. The metric is only explicitly known for $\Gamma=\mathbb{Z}_m$ in which case it is given by the Gibbons-Hawking metric \cite{Gibbons:1979zt} which is a generalization of the Eguchi-Hanson metric \cite{EGUCHI1978249}. For the D- and E-series, the metric is known to exist \cite{Kronheimer:1989zs} as a hyper-K\"ahler resolution of a quotient singularity $\mathbb{C}^{2}/\Gamma$, but its explicit form has yet to be established.

Likewise, ALF metrics are also diffeomorphic to $\mathbb{R}^{+}\times \text{S}^3/\Gamma$. However, their metric instead asymptotically approaches the metric on $(\mathbb{R}^{3} \times \text{S}^{1})/\Gamma$ . In fact 
they are only known to exist for the A- and D-series and not for the E-series of the ADE-classification \cite{Cherkis:1998hi}. In the case of the A-series they are the Taub-NUT spaces with NUT charge $m$
\cite{Hawking:1976jb} constructed as a circle fibration over a base $\mathbb{R}^{3}$. At the location of the centers the circle fiber shrinks to zero size whereas at infinity the radius becomes constant. In the 
case of the D-series the metric is only known explicitly asymptotically but an implicit construction of the full metric may be found in \cite{Cherkis:2003wk}, see also \cite{Dancer:1992kn} for the metric with quotient $\mathbb{D}_{1}$. Sufficiently far from the center, the metric can be approximated by a Taub-NUT space with $2m$ centers, plus a contribution to the centers with a negative mass parameter, subject to a quotient by $\mathbb{Z}_2$ which acts by reflecting the coordinates on both $\mathbb{R}^3$ and $\text{S}^1$ \cite{Sen:1997kz}. In the black string solutions we always take the ALE and ALF spaces in their singular limit: i.e. with a $\mathbb{C}^2/\Gamma$ singularity at the center. The black string metric however, has a smooth near horizon limit. Note that ALF spaces admit a circle at infinity whilst ALE spaces lack such a circle. This will play a crucial role in the calculation of one-loop corrections to the current levels in section \ref{sec:one-loop}.

\begin{table}[h]
\begin{center}
\begin{tabular}{lcc}
\specialrule{.09em}{0.05em}{0em}
$\Gamma\subset \text{SU}(2)$      \qquad     &     \qquad      $|\Gamma|$     \qquad    &     \qquad   \text{singularity type}\\\specialrule{.07em}{0.05em}{0em}
cyclic group $\mathbb{Z}_m$ \qquad   &    \qquad        $m$                  \qquad                        &   \qquad   $A_{m-1}$\\[0.05cm]
binary dihedral  $\mathbb{D}^*_{m}$  \qquad  &    \qquad    $4m$              \qquad        &   \qquad   $D_{m+2}$\\[0.05cm]
binary tetrahedral  $\mathbb{T}^*$    \qquad       &    \qquad   $24$        \qquad            &   \qquad     $E_6$\\[0.05cm]
binary octahedral  $\mathbb{O}^*$    \qquad       &    \qquad   $48$           \qquad         &    \qquad    $E_7$\\[0.05cm]
binary icosahedral  $\mathbb{I}^*$      \qquad     &    \qquad   $120$         \qquad           &   \qquad     $E_8$\\
\specialrule{.09em}{.05em}{-1em}
\end{tabular}
\end{center}
\caption{The freely acting discrete subgroups of $SU(2)$.} 
\label{ADE}
\end{table}

Both of the setups outlined above lead to six-dimensional black strings. It is known \cite{Couzens:2017way}, that the near horizon geometry of such a black string with $\mathcal{N}=(0,4)$ supersymmetry must be
\begin{equation}
\text{AdS}_3 \times \text{S}^3/\Gamma \, .\lab{near-hor}
\end{equation}
For the A-series these black strings have a $U(1)_{L} \times SU(2)_{R}$ isometry group of the transverse space and may be spinning. For the D- and E-series the isometry group of the transverse space is $SU(2)_R$ and the strings are static. We may reinterpret these black string solutions as five-dimensional black holes by compactifying on the S$^{1}$ wrapped by the D3 common to both setups. These 5d black holes have a near horizon geometry AdS$_{2}\times$S$^{3}/\Gamma$. For ALF spaces one may perform a further compactification on the asymptotic circle to obtain four-dimensional black holes. The entropy of these black holes follows from the left central charge and level (for the A-series) via  Cardy's formula. 

In contrast to the $\mathbb{R}^{4}$ case, where the SCFT has $\mathcal{N}=(4,4)$ supersymmetry, the black string solutions with either ALE or ALF transverse spaces are dual to chiral $\mathcal{N}=(0,4)$ SCFTs. This is manifest in the different current algebras admitted in the latter cases compared to the $\mathbb{R}^{4}$ one. The current algebras of the dual 2d SCFTs are realized holographically by the isometries of the solution. Therefore the A-series SCFT has a $U(1)_{L}\times SU(2)_{R}$ current algebra. In the D and E cases the $U(1)_{L}$ is broken, implying the SCFT has only an $SU(2)_{R}$ current algebra. To each current there is an associated level, denoted by $k_{L,R}$, and in the case of the R-symmetry this uniquely determines the right-moving central charge. In all cases the $SU(2)_{R}$ is identified with the $SU(2)_{r}$ R-symmetry of the small superconformal algebra of a 2d $\mathcal{N}=(0,4)$ SCFT and supersymmetry implies that the right-moving central charge and right-moving level are related via $c_R=6 k_R$.\footnote{This is a subtle point and technically only applies when one decouples the contributions of the center of mass modes, see \cite{Dabholkar:2010rm}. However as we only compute the levels and $c_L-c_R$ on both sides this will not matter for the matching between the microscopics and the macroscopics. We discuss this issue in more detail in section \ref{sec: ale micro}.}


\section{Macroscopics of 6d strings}\label{sec:macro}

In this section we compute the central charges and levels corresponding
to the setups described in the previous section using six-dimensional $\mathcal{N}=(2,0)$ supergravity \cite{Townsend:1983xt,Romans:1986er,Riccioni:1997np}. As we shall review below, compactifying type
IIB supergravity on K3 results in a gravity multiplet coupled to $21$
tensor multiplets \cite{Townsend:1983xt,Romans:1986er}. The macroscopic configuration is that of a black string solution in 6d with near horizon geometry AdS$_3 \times \mathrm{S}^{3}/\Gamma$ which asymptotically approaches 
$\mathbb{R}^{1,1}\times M_{\Gamma \infty},$ with $M_{\Gamma \infty}$ denoting
the asymptotic geometry of the ALE/ALF space $M_\Gamma$. In principle one could compute the central charges and levels by dimensionally reducing the six-dimensional action on the compact spherical part of the near horizon geometry in order to obtain an effective action on AdS$_3$ \cite{Kraus:2005vz,Kraus:2005zm,Hansen:2006wu}. Using the AdS/CFT dictionary the central charges and levels then correspond to the coefficients of Chern--Simons terms in the dimensionally reduced action. Concretely, $k_{L,R}$ correspond (when the former is present) to the coefficients of the $U(1)_L$ (when present) and $SU(2)_R$ Chern--Simons terms respectively, whilst $c_L-c_R$ gives the coefficient of the gravitational Chern--Simons term.

However, it was shown in \cite{Dabholkar:2010rm,Grimm:2018weo} that this is not the full story and does \emph{not} generically reproduce the correct microscopic results. Instead it was shown that one must perform the reduction of the spherical part at spatial infinity. It is known that black holes can have degrees of freedom living outside the horizon, known as hair, which contribute to the microscopic degeneracy \cite{Banerjee:2009uk,Jatkar:2009yd}. It was shown in \cite{Dabholkar:2010rm} that by performing the reduction at asymptotic infinity one includes the contributions of the hair. Of course one could also perform the reduction in the near horizon and in addition compute the contributions of the hair, the sum of which will give the result at asymptotic infinity. Clearly the near horizon analysis by itself is insufficient as the distinction between the ALE and ALF cases is not visible there.

In addition to performing the reduction at infinity one is moreover required to include contributions from one-loop Chern--Simons terms arising from integrating out massive Kaluza-Klein (KK) modes \cite{Grimm:2018weo}. For the setups considered here, the classical contributions are the same whether they are computed in the near horizon geometry or at asymptotic infinity, the quantum corrections, on the other hand do depend on this. For this reason we shall perform computations at asymptotic infinity, reducing the action using the asymptotic solution. Having said this, it is sometimes more convenient to perform the computation in the near horizon geometry when it is known that the asymptotic computation will agree with the near horizon one, i.e. when the contributions of the hair vanish.

We begin this section with a review of six-dimensional $\mathcal{N}=(2,0)$ supergravity arising from type IIB supergravity compactified on K3. We then compute the classical and quantum corrections to the central charges and levels by first reducing the six-dimensional action to three dimensions and then including the one-loop contributions. We conclude the section with a summary of the results obtained therein in anticipation of comparing with the microscopic computations in the subsequent section.

\subsection{Six-dimensional supergravity from type IIB on K3}\label{sec:6dsugra}

The massless bosonic field content of type IIB supergravity includes a metric, a complex scalar, two real two-forms and a four-form with self-dual field strength. The cohomology of the manifold K3 admits one scalar, three self-dual two-forms, nineteen anti-self-dual two-forms and one four-form. As usual the reduction to the massless 6d sector follows by expanding the various type IIB supergravity fields in terms of the generators of the cohomology of K3. Reducing the four-form leads to one scalar, three self-dual rank two tensors and nineteen anti-self-dual rank two tensors. The two 10d two-forms each lead to twenty-two scalars, one self-dual rank two tensor and one anti-self-dual rank two tensor in 6d. The complex scalar trivially reduces to two real scalars, whilst the metric leads to the graviton in 6d and fifty-eight scalars. In total there are 105 scalars, 21 anti-self-dual rank two tensors, five self-dual rank two tensors and the graviton. 

Having determined the massless bosonic fields in 6d we may now rearrange them into 6d $\mathcal{N}=(2,0)$ multiplets. The various representations, labeled by the spins  $(j_{1},j_{2})$ of the little group $SO(4)\simeq SU(2)_{1}\times SU(2)_{2}$, are:
\begin{itemize}
\item one gravity multiplet: $(1,1)\oplus4(1,\frac{1}{2})\oplus5(1,0)$,
containing one graviton, 2 left-handed gravitinos and five self-dual rank
two tensors,
\item $n_{T}$ tensor multiplets: $(0,1)\oplus4(0,\frac{1}{2})\oplus5(0,0)$, 
containing one anti-self-dual rank two tensor, two right-handed tensorinos
and 5 real scalars.
\end{itemize}
As sketched above, the compactification on K3 results in a theory coupled to $n_{T}=21$ tensor fields. This is the exact number of tensors necessary for the theory to be anomaly free and is in fact the 
unique anomaly-free six-dimensional $\mathcal{N}=(2,0)$
theory \cite{Townsend:1983xt}. Uniqueness follows from the fact that the only possible matter multiplet in chiral $\mathcal{N}=(2,0)$ supergravity is the tensor multiplet and that the cancellation of gravitational anomalies requires $n_{T}=21$.

To perform computations it is advantageous to collectively denote the tensors in the gravity- and tensor multiplets by $\hat{B}^{\alpha},$ with $\alpha=1,...,26$, and their corresponding (self-dual) field strengths as $\hat{G}^{\alpha}=\mathrm{d}\hat{B}^{\alpha}$.  
The 105 scalars in the tensor multiplets parametrize the coset 
\begin{equation}
\mathcal{M}_{\mathrm{tensor}}=\frac{SO(5,21)}{SO(5)\times SO(21)}\,.
\end{equation}
It turns out that in order to parametrize the scalar manifold it is more convenient to use the 130 scalars $\hat{\jmath}_{k}^{\alpha}$,
$k=1,...,5,$ satisfying the 25 constraints
\begin{equation}
\Omega_{\alpha\beta}\hat{\jmath}_{k}^{\alpha}\hat{\jmath}_{l}^{\beta}=\delta_{kl}\,.
\end{equation}
Here $(\Omega_{\alpha\beta})$ is the $SO(5,21)$ invariant constant metric with mostly minus signature. It may be used to define a positive definite metric via
\begin{equation}
g_{\alpha\beta}=2\hat{\jmath}_{k\alpha}\hat{\jmath}_{\beta}^{k}-\Omega_{\alpha\beta}\,,\qquad\hat{\jmath}_{k\alpha}\equiv\Omega_{\alpha\beta}\hat{\jmath}_{k}^{\beta}\,,\qquad\hat{\jmath}^{k\alpha}\equiv\delta^{kl}\hat{\jmath}_{l}^{\alpha}\,.
\end{equation}
With this formalism the bosonic part of the pseudo-action is given by \cite{Riccioni:1997np}\footnote{Note that our conventions differ with respect to \cite{Riccioni:1997np},
i.e. $\hat{G}_{\mathrm{there}}^{\alpha}=\tfrac{1}{2}\hat{G}_{\mathrm{here}}^{\alpha}.$} 
\begin{equation}
S^{(6)}=\frac{1}{(2\pi)^{3}}\int_{\mathcal{M}_{6}}\Big[\tfrac{1}{2}\hat{R}\hat{\ast}1-\tfrac{1}{4}g_{\alpha\beta}\hat{G}^{\alpha}\wedge\hat{\ast}\hat{G}^{\beta}-\tfrac{1}{2}(\hat{\jmath}_{k\alpha}\hat{\jmath}_{\beta}^{k}-\Omega_{\alpha\beta})\mathrm{d}\hat{\jmath}_{l}^{\alpha}\wedge\hat{\ast}\mathrm{d}\hat{\jmath}^{l\beta}\Big]\,,
\end{equation}
where we chose conventions such that $\kappa_{6}^{2}=(2\pi)^{3}.$
This is a pseudo-action since the self-duality constraints for the
tensors, 
\begin{equation}\label{self-duality condition}
g_{\alpha\beta}\hat{\ast}\hat{G}^{\beta}=\Omega_{\alpha\beta}\hat{G}^{\beta},
\end{equation}
do not follow from the action and must be imposed by hand at the level
of the equations of motion.

\subsection{Black string solutions}

Recall that we are interested in black strings arising from wrapping a D3-brane on a curve inside K3. As such we cannot turn on tensor fields arising from the reduction of the two 10d two-forms, each providing a self-dual and anti-self-dual tensor field. Moreover, preservation of supersymmetry implies that the Poincar\'e dual of the wrapped curve is in $H^{1,1}(K3)$ \cite{Cvetic:1997yf,Gauntlett:2007ph,Couzens:2017way}. This implies that the six-dimensional string is only charged under tensors arising from the reduction of the four-form along two-forms in $H^{1,1}(K3)$. Consequently we may only have non-vanishing values for one self-dual tensor and 19 anti-self-dual tensors. Note that these are exactly the tensors we find in $\mathcal{N}=(1,0)$
supergravity coupled to 19 tensor multiplets. This theory has black string solutions with a transverse hyper-K\"ahler
space \cite{Lam:2018jln, Gutowski:2003rg}. These solutions may be embedded into the $\mathcal{N}=(2,0)$
theory by setting the scalars $\hat{\jmath}_{k}^{\alpha}=\delta_{k}^{27-\alpha}$
for $k=2,3,4,5$ and $\hat{\jmath}_{1}^{\alpha}=0$ for $\alpha=21,...,26.$\footnote{For later simplicity we have taken a non-canonical choice for $\Omega$. We take the first diagonal entry to be $+$, the next 21 $-$ and the last four entries to be $+$.}
Therefore, from now on, $\alpha=1,...,20,$ and the scalars $\hat{\jmath}^{\alpha}\equiv\hat{\jmath}_{1}^{\alpha}$
satisfy
\begin{equation}
\Omega_{\alpha\beta}\hat{\jmath}^{\alpha}\hat{\jmath}^{\beta}=1\,,
\end{equation}
where $(\Omega_{\alpha\beta})$ is the (canonical) $SO(1,19)$ invariant constant
metric with mostly minus signature. It can be identified with the
metric on $H^{1,1}(\mathrm{K3})$
\begin{equation}
\Omega_{\alpha\beta}=\eta_{\alpha\beta}=\int_{\mathrm{K3}}\omega_{\alpha}\wedge\omega_{\beta}\,.
\end{equation}
 Black string solutions have a metric of the form \cite{Lam:2018jln,Gutowski:2003rg}
\begin{equation}
\mathrm{d}\hat{s}_{6}^{2}=2H^{-1}\left(\mathrm{d}u+\beta\right)\Big(\mathrm{d}v+\omega+\frac{1}{2}\mathcal{F}\left(\mathrm{d}u+\beta\right)\Big)+H\mathrm{d}s^{2}(M_{\Gamma})\,,\label{eq:black string solution}
\end{equation}
where $\mathrm{d}s^2(M_{\Gamma})$ is either the ALE or ALF hyper-K\"ahler
metric corresponding to the finite group $\Gamma.$ The one-forms
$\omega,$ $\beta$ are defined on $M_{\Gamma}$ and, like the functions
$H$ and $\mathcal{F},$ are independent of $u$ and $v$. They
satisfy certain equations \cite{Lam:2018jln} that in principle can
be solved once an explicit expression for the hyper-K\"ahler metric
is given. 

Note that $\partial_{u}$ is a Killing vector
of the metric (\ref{eq:black string solution}) and taking it to be spacelike
we find a metric for a black string wound in the $u$-direction.
The near horizon geometry of this string is $\mathrm{AdS}{}_{3}\times\mathrm{S}^{3}/\Gamma$ \cite{Couzens:2017way}
with asymptotics given by (\ref{eq:black string solution})
up to replacing $\mathrm{d}s^2(M_\Gamma)$ by its asymptotic metric
$\mathrm{d}s^2(M_{\Gamma \infty}).$ The asymptotic space has a finite covering and its metric  approaches a $\Gamma$-quotient of the following metric sufficiently fast:
\begin{eqnarray} \label{metrics asymptotics}
\mathrm{d}s^2(M_{\infty}) & = & \begin{cases}
\mathrm{d}r^{2}+r^{2}(\sigma_{1}^{2}+\sigma_{2}^{2}+\sigma_{3}^{2})\, & M_\Gamma \:\mathrm{is\:ALE}\,,
\vspace{.2cm}\\
\mathrm{d}r^{2}+r^{2}(\sigma_{1}^{2}+\sigma_{2}^{2})+\sigma_{3}^{2} & M_\Gamma \:\mathrm{is\:ALF}\,.
\end{cases}
\end{eqnarray}
Here $\sigma_{i}$ are the left-invariant forms on $\mathrm{S}^{3}.$
We use the covering space to treat all quotients simultaneously.
The charges of the black string follow from integrating the three-forms
$\hat{G}_{\Gamma}^{\alpha}$ in the quotient space over the spherical
part of the metric (given by the $\sigma_{i}$ part). We choose the normalization
\begin{equation}
\int_{M_{\Gamma}^{\mathrm{sph}}}\hat{G}_{\Gamma}^{\alpha}=\frac{1}{|\Gamma|}\int_{M^{\mathrm{sph}}}\hat{G}^{\alpha}=-(2\pi)^2Q^{\alpha}\,,
\end{equation}
by using the quotient map $\pi:\;M\rightarrow M_{\Gamma},$ $\hat{G}^{\alpha}=\pi^{*}\hat{G}_{\Gamma}^{\alpha}.$
These charges are related to the microscopic charges $q^{\alpha}$
via 
\begin{equation}
Q^{\alpha}=q^{\alpha}\,.
\end{equation}
This identification may be proven in a similar manner as in \cite{Grimm:2018weo}. Note that contrary to \cite{Grimm:2018weo} we do not find a shift in the charge. This is because the action in the K3 case does not include a higher-derivative term  $\propto \int_{\mathcal{M}_6}\hat{B}^\alpha \wedge \mathrm{tr} \, \hat{\mathcal{R}} \wedge \hat{\mathcal{R}}$ present in the setup considered in \cite{Grimm:2018weo} which leads to such a shift.


\subsection{Classical contribution for ALE and ALF transverse spaces \label{subsec:Classical-contribution}}

In this section we determine the classical contribution to the central
charges and levels. Here `classical' refers to the contribution
that is obtained by reducing the six-dimensional $\mathcal{N}=(2,0)$ pseudo-action
to three dimensions along the compact space. The contributions to the central charges and levels are given by the coefficients
of 3d Chern--Simons terms. We therefore only need to consider the
part of the pseudo-action that can give rise to terms of this form, specifically the term:
\begin{equation}
-\frac{1}{32\pi{}^{3}}\int_{\mathcal{M}_{6}}g_{\alpha\beta}\hat{G}^{\alpha}\wedge\hat{\ast}\hat{G}^{\beta}\subset S^{(6)}\,.\label{relevant piece of the action}
\end{equation}
The reduction for the black strings with either
ALE or ALF transverse space will be the same and therefore we treat them concurrently.

To perform the reduction one gauges the isometries of S$^3/\Gamma$ by introducing gauge fields for these symmetries. This requires a modification of the three-form flux and we take the ansatz for $\hat{G}^{\alpha}$, the pull back of the three-form
on the quotient space, to be \cite{Hansen:2006wu,Dabholkar:2010rm}\footnote{Strictly the expression \eqref{ansatz three-form} only satisfies the self-duality constraint after setting the gauge fields to zero. However this is sufficient as the gauge fields are treated as fluctuations around the background.}
\begin{equation}
\hat{G}^{\alpha}=-Q^{\alpha}\big[(2\pi)^{2}|\Gamma|\big(e_{3}-\chi_{3}\big)+\omega\big(\mathcal{M}_3\big)\big]\, .\label{ansatz three-form}
\end{equation}
The three-dimensional space $\mathcal{M}_3$ is the non-spherical part of the covering space and $\omega\big(\mathcal{M}_3\big)$ is a three-form on $\mathcal{M}_{3}$ whose form we do not specify. The three-form $e_{3}$ appearing in (\ref{ansatz three-form})
is invariant under the isometries of the spacetime, integrates to unity over the spherical part 
\begin{equation}
\int_{M^{\mathrm{sph}}}e_{3}=1\, ,
\end{equation}
and has exterior derivative 
\begin{equation}
\mathrm{d}e_{3}=\begin{cases}
\frac{1}{16\pi^{2}}F_{L}\wedge F_{L}+\frac{1}{8\pi^{2}}\mathrm{tr}\,F_{R}\wedge F_{R} & \Gamma=\mathbb{Z}_{m}\,,
\vspace{.2cm}\\
\frac{1}{8\pi^{2}}\mathrm{tr}\,F_{R}\wedge F_{R} & \Gamma\neq\mathbb{Z}_{m}\,.
\end{cases}
\end{equation}
An explicit form for this three-form can be obtained from \cite{Hansen:2006wu}. 
One needs to set the gauge fields corresponding
to the broken symmetries to zero, that is only keeping gauge fields corresponding
to $U(1)_{L}\times SU(2)_{R}$ for $\Gamma=\mathbb{Z}_{m}$ and those corresponding to $SU(2)_{R}$ for other $\Gamma$.
The three-form $\chi_{3}$ in (\ref{ansatz three-form})
is given by
\begin{equation}
\chi_{3}=\begin{cases}
\frac{1}{16\pi^{2}}A_{L}\wedge F_{L}+\frac{1}{8\pi^{2}}\mathrm{tr}\big(A_{R}\wedge\mathrm{d}A_{R}+\frac{2}{3}A_{R}^{3}\big) & \Gamma=\mathbb{Z}_{m}\,,
\vspace{.2cm}\\
\frac{1}{8\pi^{2}}\mathrm{tr}\big(A_{R}\wedge\mathrm{d}A_{R}+\frac{2}{3}A_{R}^{3}\big) & \Gamma\neq\mathbb{Z}_{m}\,,
\end{cases}
\end{equation}
and is included to ensure that $\hat{G}^{\alpha}$ satisfies its Bianchi identity.

We are now ready to determine the contribution of (\ref{relevant piece of the action})
to the central charges and levels. We calculate it by determining
the gauge variation of the reduced action under either $U(1)_{L}\times SU(2)_{R}$
gauge transformations for $\Gamma=\mathbb{Z}_{m}$ or $SU(2)_{R}$
gauge transformations for $\Gamma\neq\mathbb{Z}_{m}$. We parametrize
the gauge transformations by $\Lambda$ and determine the variation
of the three-dimensional action by reducing the variation of the six-dimensional
action on $M_{\Gamma}^{\mathrm{sph}}.$ By construction $e_{3}$ is
invariant under all gauge transformations, consequently only the variation of $\chi_{3}$ contributes
to the variation of the six-dimensional action. We find\footnote{In our conventions $\int_{\mathcal{M}_{6}}=\int_{\mathcal{M}_{\Gamma3}}\cdot\int_{M_{\Gamma}^{\mathrm{sph}}},$
where $\mathcal{M}_{\Gamma3}$ is the three-dimensional non-spherical part
of the quotient space.}
\begin{eqnarray}
\delta_{\Lambda}\mathcal{L}_{\mathrm{CS}}\ast_{3}1 & = & -\frac{1}{16\pi^{3}}\int_{M_{\Gamma}^{\mathrm{sph}}}g_{\alpha\beta}\delta_{\Lambda}\hat{G}_{\Gamma}^{\alpha}\wedge\hat{\ast}\hat{G}_{\Gamma}^{\beta}=-\frac{1}{16\pi^{3}|\Gamma|}\int_{M^{\mathrm{sph}}}g_{\alpha\beta}\delta_{\Lambda}\hat{G}^{\alpha}\wedge\hat{\ast}\hat{G}^{\beta}\nonumber \\
 & = & \pi|\Gamma|\eta_{\alpha\beta}Q^{\alpha}Q^{\beta}\int_{M^{\mathrm{sph}}}\delta_{\Lambda}\chi_{3}\wedge e_{3}=\pi|\Gamma|\eta_{\alpha\beta}Q^{\alpha}Q^{\beta}\delta_{\Lambda}\chi_{3}\,,\label{variation 6d action}
\end{eqnarray}
where the third equality sign follows by using the (anti-)self-duality
condition (\ref{self-duality condition}). This gives the gauge variation of the three-dimensional action which from general considerations is given by
\begin{eqnarray}
S_{\mathrm{CS}} & = & \pi|\Gamma|\eta_{\alpha\beta}Q^{\alpha}Q^{\beta}\int_{\mathcal{M}_{\Gamma 3}}\chi_{3}\nonumber \\
 & = & \begin{cases}
\frac{k_{L}^{\mathrm{class}}}{8\pi}\int_{\mathcal{M}_{\Gamma 3}}A_{L}\wedge F_{L}+\frac{k_{R}^{\mathrm{class}}}{4\pi}\int_{\mathcal{M}_{\Gamma 3}}\mathrm{tr}\big(A_{R}\wedge\mathrm{d}A_{R}+\frac{2}{3}A_{R}^{3}\big) & \Gamma=\mathbb{Z}_{m}\,,
\vspace{.2cm}\\
\frac{k_{R}^{\mathrm{class}}}{4\pi}\int_{\mathcal{M}_{\Gamma 3}}\mathrm{tr}\big(A_{R}\wedge\mathrm{d}A_{R}+\frac{2}{3}A_{R}^{3}\big) & \Gamma\neq\mathbb{Z}_{m}\,.
\end{cases}
\end{eqnarray}
Comparing the two expressions we thus find for the central charges and levels
\begin{eqnarray}
k_{L}^{\mathrm{class}} & = & \frac{1}{2}|\Gamma|\eta_{\alpha\beta}Q^{\alpha}Q^{\beta} = \frac{1}{2}|\Gamma|C\cdot C\quad\mathrm{only\:for\:\Gamma}=\mathbb{Z}_{m}\,,\nonumber \\
k_{R}^{\mathrm{class}}&=&\frac{1}{2}|\Gamma|C\cdot C\, ,\label{classical contribution central charges and levels} \\
c_{L}^{\mathrm{class}} & = & c_{R}^{\mathrm{class}}\, ,\nonumber
\end{eqnarray}
where we have introduced the notation
\begin{equation}
C\cdot C= \int_{\text{K}3} C\wedge C = q^{\alpha} \eta_{\alpha \beta} q^{\beta}\,.
\end{equation}
In writing \eqref{classical contribution central charges and levels} we have used that $c_L^{\mathrm{class}}-c_R^{\mathrm{class}}=0$ due to the absence of a 3d gravitational Chern--Simons term. It is tempting to set $c_{R}=6 k_{R}$ using the field theory result, however as we will explain in section \ref{sec: ale micro} this is only true modulo center of mass modes. 

We note that if we had performed the reduction in the near horizon geometry we would have obtained exactly the same results as in (\ref{classical contribution central charges and levels}).
This is a quirk of the current setup, in general a difference in the two reductions is possible when the supergravity action contains a higher derivative term, see \cite{Grimm:2018weo}, but as such a term is missing here the results agree.
When $M_{\Gamma}$ is ALE \eqref{classical contribution central charges and levels} gives the full answer. On the other hand for ALF transverse space one must also include one-loop contributions originating from integrating out massive Kaluza-Klein modes. The content of the next section is devoted to the calculation of these one-loop contributions and showing that they only contribute in the ALF case and not the ALE one.


\subsection{Quantum contributions for ALF transverse spaces}\label{sec:one-loop}

In \cite{Grimm:2018weo} it was shown that including contributions from one-loop Chern--Simons terms
is essential for reproducing the correct central charges and levels.
These terms arise from integrating out massive Kaluza-Klein (KK) modes
running in the loops of the relevant two-point functions. The relevant
KK modes come from the chiral fields in six dimensions, i.e. the six-dimensional
gravitinos, the spin-$\frac{1}{2}$ fermions in the tensor multiplets
and the (anti-)self-dual two-forms. After dimensional reduction to three dimensions these
fields lead to massive spin-$\frac{3}{2},$ spin-$\frac{1}{2}$ and
chiral vector fields. For simplicity we shall perform the calculation of the one-loop Chern--Simons terms in the AdS$_3\times \mathrm{S}^{3}/\Gamma$ near horizon geometry. Given our previous comments this may seem contradictory. However, we shall argue that for ALF transverse spaces the near horizon computation matches the analogous computation at infinity. Further we will explain why for ALE transverse spaces the quantum contributions vanish when dimensionally reducing in the asymptotic geometry. 

Let us end this introduction by outlining the strategy we will follow in the remainder of the section to compute the quantum corrections. We begin by giving the relevant KK spectrum for $\mathcal{N}=(2,0)$ supergravity on AdS$_3\times \mathrm{S}^3$ as determined in \cite{Deger:1998nm,deBoer:1998kjm}. In particular, for each mode we give the three-dimensional Lorentz representation, the $\mathcal{\mathfrak{so}}(4)=\mathfrak{su}(2)_{L}\oplus\mathfrak{su}(2)_{R}$ representation following from the isometries of the three-sphere, and the sign of the mass. We shall then determine which modes are projected out under the action of $\Gamma= \mathbb{Z}_{m}$ and $\Gamma=\mathbb{D}^{*}_{m}$, thereby determining the KK-spectrum after the reduction on AdS$_3\times \mathrm{S}^{3}/\Gamma$. Having determined the spectrum we may use the contribution of each mode to the three-dimensional $\mathfrak{u}(1)_{L}$ (only when $\Gamma=\mathbb{Z}_{m}$), $\mathfrak{su}(2)_{R}$ and gravitational Chern--Simons terms that were determined in \cite{Grimm:2018weo} using the Atiyah-Patodi-Singer index theorem. Finally we sum the individual contributions over all modes. It is necessary to regularize the summations using zeta-function regularization in order to obtain a finite result for the quantum corrections to the central charges and levels. Adding these to the classical part determined in section \ref{subsec:Classical-contribution} we obtain the microscopic data up to order $\mathcal{O}(1)$ in the charges.

\subsubsection{Kaluza-Klein spectrum \label{subsec:Kaluza-Klein-spectrum}}

The six-dimensional fields giving rise to massive three-dimensional
chiral modes are the 2 gravitinos, $2n_{T}=42$ tensorinos, $5$
self-dual rank two tensors and $n_{T}=21$ anti-self-dual rank two tensors.
The gravitinos and tensorinos are given by two Weyl fermions
obeying a symplectic-Majorana condition and the tensors are subject
to a reality condition. The gravitinos give rise to three-dimensional
spin-$\frac{3}{2}$ and spin-$\frac{1}{2}$ particles, the tensorinos
to spin-$\frac{1}{2}$ particles and the (anti-)self-dual tensors
to chiral vector fields. The KK spectrum of AdS$_3\times$S$^3$ was computed in \cite{Deger:1998nm,deBoer:1998kjm}. Initially we will not take into account the symplectic-Majorana nor reality condition, instead imposing these later. The representations are labeled in terms of $\mathfrak{su}(2)_{L}\oplus\mathfrak{su}(2)_{R}$ and the sign of the mass, $\mathrm{sgn}(M)$, as $(j_{L},j_{R})^{\mathrm{sgn}(M)}$\,\footnote{We use notation similar to \cite{deBoer:1998kjm}. The same spectrum
in \cite{Deger:1998nm} is listed in terms of the highest weight vector
$(l_{1},l_{2})$ of $\mathfrak{so}(4).$ The relation with our notation
is given by $l_{1}=j_{L}+j_{R},$ $l_{2}=j_{L}-j_{R}.$ }.
The representations are:
\begin{itemize}
\item Spin-$\frac{3}{2}$:
\[
4\bigoplus_{j_{L}=\frac{1}{2}}^{\infty}\big(j_{L},j_{L}\pm\tfrac{1}{2}\big)^{\mp}.
\]
\item Spin-$\frac{1}{2}$:
\begin{eqnarray*}
4\bigoplus_{j_{L}=\frac{3}{2}}^{\infty}\big(j_{L},j_{L}\pm\tfrac{3}{2}\big)^{\mp} & \oplus & 4\bigoplus_{j_{L}=0}^{1}\big(j_{L},j_{L}+\tfrac{3}{2}\big)^{-}\oplus 4 \bigoplus_{j_{L}=1}^{\infty}\big(j_{L},j_{L}\pm\tfrac{1}{2}\big)^{\pm}\oplus 4 \big(\tfrac{1}{2},1\big)^{+}\\
\oplus \, 4 \big(0,\tfrac{1}{2}\big)^{+} & \oplus & 84 \bigoplus_{j_{L}=\frac{1}{2}}^{\infty}\big(j_{L},j_{L}\pm\tfrac{1}{2}\big)^{\pm}\oplus 84 \big(0,\tfrac{1}{2}\big)^{+}.
\end{eqnarray*}
\item Chiral vectors:
\[
5\bigoplus_{j_{L}=1}^{\infty}\big(j_{L},j_{L}\pm1\big)^{\mp}\oplus 5 \big(\tfrac{1}{2},\tfrac{3}{2}\big)^{-}\oplus 5 \big(0,1\big)^{-}\oplus 21 \bigoplus_{j_{L}=1}^{\infty}\big(j_{L},j_{L}\pm1\big)^{\pm}\oplus 21 \big(\tfrac{1}{2},\tfrac{3}{2}\big)^{+}\oplus 21 \big(0,1\big)^{+}.
\]
\end{itemize}
The notation
\begin{equation}
\big(j_{L},j_{L}\pm\tfrac{1}{2}\big)^{\mp}=\big(j_{L},j_{L}+\tfrac{1}{2}\big)^{-}\oplus \big(j_{L},j_{L}-\tfrac{1}{2}\big)^{+}
\end{equation}
symbolizes two towers of KK modes. Furthermore, the three-dimensional fermions
in the above spectrum are Dirac spinors and the chiral vectors are
complex. We must apply the six-dimensional symplectic-Majorana
and reality conditions. Denoting the eigenvalues of the generators
of $\mathfrak{u}(1)_{L}\subset\mathfrak{su}(2)_{L}$ and $\mathfrak{u}(1)_{R}\subset\mathfrak{su}(2)_{R}$
by $j_{L}^{3}$ and $j_{R}^{3}$ respectively, these conditions imply
that modes with $j_{L}^{3}$ and $j_{R}^{3}$ get mapped to modes
with $-j_{L}^{3}$ and $-j_{R}^{3}$ respectively \cite{Ishiki:2006rt}. As a result we need only sum over modes with $j_{R}^{3}\geq0$ or $j_{L}^{3}\geq0$.
To determine the KK spectrum on S$^3/\Gamma$ we determine which states are invariant under the action of $\Gamma$ for $\Gamma=\mathbb{Z}_{m}$ and $\Gamma= \mathbb{D}^{*}_{m}$. The full details are presented in appendix \ref{sec:Action An and Dn on KK spectrum}. For $\Gamma=\mathbb{Z}_{m}$ the invariant states satisfy $j_{L}^{3}=\frac{1}{2}mk$ with $k \in \mathbb{Z}$, the symplectic-Majorana and reality conditions further refine this to $k\in \mathbb{Z}_{\geq0}$. For $\mathbb{D}^{*}_{m}$ the invariant representations have $j_{L}^{3}= m k$ with $k\in \mathbb{Z}_{\geq0}$. The reality conditions imply in addition that we only keep states with $j^{3}_{R}\geq0$.

At asymptotic infinity the reduction is not performed on S$^3/\Gamma$ but on the compact part of the quotient of  \eqref{metrics asymptotics}. It is therefore not a priori clear that the KK spectrum is the same as the one given above. For $\Gamma=\mathbb{Z}_{m}$ the reduction is on a squashed Lens space, where the radius of the two-sphere is taken to infinity. The Hopf circle still has finite radius and therefore the expectation is that the representation content of the KK spectrum will be the same as for S$^3/\mathbb{Z}_{m}$. We shall assume that this expectation is met and moreover that the signs of the masses do not change. For $\Gamma=\mathbb{D}^{*}_{m}$ there is still a two-sphere in the asymptotic geometry whose radius tends to infinity. However in contrast to the A-series there is no longer a circle, instead there is a segment with finite length, see for example \cite{Sen:1997kz}. Again the expectation is that we can use the previously determined KK spectrum.

This discussion sheds light on why we do not have to include one-loop Chern--Simons terms for a transverse ALE space. From \eqref{metrics asymptotics} it is easy to see that there is no compact space at asymptotic infinity on which one can perform a KK reduction, ergo there are no quantum corrections in the ALE case and the classical result of section \ref{subsec:Classical-contribution} is the final result.


\subsubsection{Contribution of massive KK modes to one-loop Chern--Simons terms}\label{sec:KKmodes}

In \cite{Grimm:2018weo} the contribution of each of the KK modes to the $\mathfrak{u}(1)_{L}$ (only relevant for $\Gamma=\mathbb{Z}_{m}$), $\mathfrak{su}(2)_{R}$ and gravitational Chern--Simons terms was computed. For a representation with quantum numbers $j_{L}^{3}$
and $j_{R}$ under $\mathfrak{u}(1)_{L}\oplus\mathfrak{su}(2)_{R}$
the constants $\alpha_{L},$ $\alpha_{R}$ and $\alpha_{\mathrm{grav}}$
in front of the Chern--Simons terms $\omega_{\mathrm{CS}}(A_{L}),$
$\omega_{\mathrm{CS}}(A_{R})$ and $\omega_{\mathrm{grav}}^{\mathrm{CS}}$
are given in table \ref{looptable}.\footnote{These constants do not depend on the scales of the geometry in a continuous way. The scales only appear via the masses of the Kaluza--Klein modes and these one-loop computations only depend on the masses via their sign, see for example \cite{Bonetti:2013ela} for a similar computation. Note that this again shows that the ALE case needs to be treated separately from the ALF case. Na\"ively, the ALE space can  be seen as a limit of the ALF space by decompactifying the asymptotic circle. As the masses are inversely proportional to the radius, they go to zero in this limit. However, in this limit the sign function is not defined.} The relation between the coefficients $\alpha_{L},$ $\alpha_{R},$
$\alpha_{\mathrm{grav}}$ of the Chern--Simons terms, after summing over all of the modes, to the central charges and levels are:
\begin{equation}
k_{L}=8\pi\alpha_{L}\, ,\qquad k_{R}=4\pi\alpha_{R}\, ,\qquad c_{L}-c_{R}=96\pi\alpha_{\mathrm{grav}}\,.
\end{equation}

\begin{table}[h]
\centering
{\renewcommand{\arraystretch}{1.6}
\begin{tabular}{c|c|c|c}
\specialrule{.09em}{0.05em}{-0.25em}
 & spin-$\frac{1}{2}$  & spin-$\frac{3}{2}$  &  chiral vectors\\[0.1cm]
\specialrule{.09em}{0.05em}{0em}
$\alpha_{L}$ & $\frac{1}{2}(j^3_{L})^{2}\left(2j_{R}+1\right)$ & $\frac{3}{2}(j^3_{L})^{2}\left(2j_{R}+1\right)$ & $-(j^3_{L})^{2}\left(2j_{R}+1\right)$\\[0.1cm]

$\alpha_{R}$ & $-\frac{1}{3}j_{R}\left(j_{R}+1\right)\left(2j_{R}+1\right)$ & $-j_{R}\left(j_{R}+1\right)\left(2j_{R}+1\right)$ & $\frac{2}{3}j_{R}\left(j_{R}+1\right)\left(2j_{R}+1\right)$\tabularnewline
 
$\alpha_{\mathrm{grav}}$ & $\frac{1}{48}\left(2j_{R}+1\right)$ & $-\frac{7}{16}\left(2j_{R}+1\right)$ & $\frac{1}{12}\left(2j_{R}+1\right)$\tabularnewline\Xhline{2\arrayrulewidth}
 
\end{tabular} 
}
\caption{Contributions of a single representation to the left-, right- and gravitational Chern--Simons terms. The table should be read as $\ax_I= \fr{\t{sgn}(M)}{4 \pi} \times$(entry of table). This is table 4.1 in  \cite{Grimm:2018weo}.}
\lab{looptable}
\end{table}

With these results we can now sum the contributions of table \ref{looptable} over the complete spectrum determined in section \ref{subsec:Kaluza-Klein-spectrum}. As we are summing over an infinite number of states the result will diverge and needs to be regularized. Following \cite{Grimm:2018weo} we use zeta-function regularization. The regularized summations that we need are:
\begin{eqnarray}
\sum_{n=1}^{\infty}1 & = & -\frac{1}{2}\,,\qquad \ \, \sum_{n=1}^{\infty}n=-\frac{1}{12}\,,\nonumber \\
\sum_{n=1}^{\infty}n^{2} & = & 0\, ,\qquad\ \ \ \sum_{n=1}^{\infty}n^{3}=\frac{1}{120}\,. \label{zeta-function regularization}
\end{eqnarray}

Before proceeding, a few comments regarding the regularization of the contributions from the massive KK spectrum are in order. Firstly, in the three-dimensional theory one expects that there is a UV cut-off determined by the scale at which gravity becomes strongly coupled \cite{ArkaniHamed:2005yv,Dvali:2007hz}. Regularization with this cut-off will agree with the zeta-function regularization performed here. Secondly it is only possible to apply zeta-function regularization because the higher dimensional theory is anomaly free \cite{Corvilain:2017luj}.

Let us proceed with the calculation of the quantum corrections for the A- and D-series. We shall present the final results here, leaving a more detailed exposition to appendix \ref{sec:Summation-of-3D one-loop corrections}.

\paragraph*{A-series. }

We must sum over the representations listed in section \ref{subsec:Kaluza-Klein-spectrum} with  $j_{L}^{3}=\frac{1}{2}mk$, $k \in \mathbb{Z}_{\geq0}$. We implement this by first summing over all states with $j_{L}^{3}=\frac{1}{2}mk$, that is over $j_{L}= \frac{1}{2} mk, \frac{1}{2} m k +1,...$ and then performing the sum over $k \in \mathbb{Z}_{\geq0}$, keeping in mind that we must regularize the summations using zeta-function regularization. This is performed explicitly in appendix \ref{sec:Summation-of-3D one-loop corrections}, with the final results
\begin{equation}\label{one-loop contributions Am series}
k_{L}^{\mathrm{loop}} = 0\,,\qquad
k_{R}^{\mathrm{loop}} = 2m\,, \qquad
(c_{L}-c_{R})^{\mathrm{loop}} = 12m\,,
\end{equation}
up to order $\mathcal{O}(1)$ in the charges. Adding these results to the classical contributions \eqref{classical contribution central charges and levels}, and using $|\mathbb{Z}_{m}|=m$, the central charges and levels to order $\mathcal{O}(1)$ in the charges are:
\begin{equation}
k_{L} =  \frac{1}{2}mC\cdot C\,,\qquad
k_{R} = \frac{1}{2}mC\cdot C+2m\,,\qquad
c_{L} =  c_{R}+12m\,.
\end{equation}

\paragraph*{D-series.}

As there is no left-moving current isometry in this case there is no left level, $k_{L}$, to compute. As in the A-series case we sum the contributions of the individual representations over the KK spectrum of section \ref{subsec:Kaluza-Klein-spectrum}, this time with $j_{L}^{3}=mk,~ k\in \mathbb{Z}_{\geq0}$ and $j_{R}^{3}\geq0$. This is most easily implemented by first summing over all representations with $j_{L}^{3}=mk,~ k\in \mathbb{Z}_{\geq0}$ and then imposing $j^{3}_{R}\geq0$. Comparing with the condition in the A-series it is clear that the first step may be achieved by using the A-series results \eqref{one-loop contributions Am series} and sending $m\rightarrow 2m$. We must still impose the second condition, which may be imposed \emph{effectively} by dividing the result of the first step by two\footnote{We use the word effectively as the contribution from the $j_{R}^{3}=0$ terms is half of the actual result using this method. The difference, however, is $\mathcal{O}(1)$ in the charges and therefore for our purposes may be neglected.}. The one-loop contributions up to order $\mathcal{O}(1)$ in the charges are:
\begin{equation}\label{one-loop contributions Dm series}
k_{R}^{\mathrm{loop}} = 2m\,, \qquad
(c_{L}-c_{R})^{\mathrm{loop}} = 12m\,.
\end{equation}
Adding these to \eqref{classical contribution central charges and levels} and using $|\mathbb{D}^{*}_{m}|=4m$, the final results for the central charges and right-moving level up to and including linear terms in the charges are:
\begin{equation}
k_{R} = 2mC\cdot C+2m\,, \qquad
c_{L} = c_{R}+12m\,.
\end{equation}

\subsection{Summary}

For the ease of the reader we conclude by collating the results of this section. In the subsequent section we shall compare the results obtained here with the microscopic computations conducted there. 

For the case of ALE transverse space we have shown that the only contribution is from the classical reduction of the action performed in section \ref{subsec:Classical-contribution}. The results for the central charges and levels are
\begin{eqnarray}
k_{L} & = & \frac{1}{2}|\Gamma|\eta_{\alpha\beta}Q^{\alpha}Q^{\beta}=\frac{1}{2}|\Gamma|C\cdot C\quad\qquad \text{only for $\Gamma=\mathbb{Z}_m$}\, ,\nonumber \\
k_{R}&=&\frac{1}{2}|\Gamma|C\cdot C\,,\lab{ALEmacroresults}\\
c_{L} & = & c_{R}\, , \nonumber
\end{eqnarray}
where $|\Gamma|$ may be read off of table \ref{ADE}. 

In contrast, for the case of ALF transverse space we must also include, in addition to the classical contribution of section \ref{subsec:Classical-contribution}, the quantum corrections obtained in the preceding section. For the A-series we find\footnote{Note that the central charges and levels do not follow from the results of \cite{Grimm:2018weo} by specializing to CY$_3=$K3$\times T^2$. This is due to the presence of non-trivial one-cycles that are not present in Calabi--Yau threefolds with $SU(3)$ holonomy.}
\begin{equation}\lab{Aseriesmacroresults}
k_{L} =  \frac{1}{2}mC\cdot C\,,\qquad
k_{R}  = \frac{1}{2}mC\cdot C+2m\,, \qquad
c_{L} = c_{R}+12m\,,
\end{equation}
whilst for the D-series we find:
\begin{equation}
k_{R} =  2mC\cdot C+2m\,, \qquad
c_{L} = c_{R}+12m \,.\lab{Dmacro}
\end{equation}
All results presented here are up to linear terms in the charges, we do not include any $\mathcal{O}(1)$ terms.

\section{Microscopics of 6d strings}\label{sec:microscopics}

In the previous section we have computed the central charges and levels macroscopically; the content of this section is to reproduce these results from a microscopic computation. Concretely, we shall compute the central charges and levels of the 2d $\mathcal{N}=(0,4)$ SCFTs living on the strings considered previously. For the case of transverse ALE space we utilize the results of \cite{Douglas:1996sw,Johnson:1996py} to determine the worldvolume theory on D3-branes probing an ADE singularity. By placing the resultant four-dimensional SCFTs on $\mathbb{R}^{1,1}\times C$ and performing a suitable topological twist \cite{Bershadsky:1995vm,Kapustin:2006hi, Putrov:2015jpa, Amariti:2017cyd} we obtain 2d $\mathcal{N}=(0,4)$ SCFTs, whose central charges and levels we may compute via a spectrum computation and anomaly arguments. In the ALF case it is convenient to use a dual M-theory description for the A-series. However, for the ALF D-series case such a clean microscopic setup is lacking and the matching of the microscopic results to the macroscopic ones remains an open problem.

\subsection{Transverse ALE spaces: ADE-series} \label{sec: ale micro}

We first consider the microscopic theory describing a D3-brane wrapping a curve $C$ inside K3 and probing the singular limit of an ALE space $\mathbb{C}^2/\Gamma$ transverse to the brane. The worldvolume theory of a D3-brane on $\mathbb{R}^{1,3}$ probing a transverse ADE singularity is obtained by performing a projection of $\mathcal{N}=4$ $U(|\Gamma|)$ super Yang--Mills. The resulting theories are $\mathcal{N}=2$ supersymmetric gauge theories with known Lagrangian description  \cite{Douglas:1996sw,Johnson:1996py}. Their gauge and hypermultiplet field content is summarized in table \ref{tab:Quivers-for-the}. From the table we can read off that the 4d $\mathcal{N}=2$ theories all have $n_{H}=n_{V}=|\Gamma|$ hyper- and vector multiplets. Our ultimate interest is the 2d IR theory arising from wrapping the D3-brane on a curve $C$ inside K3, preserving $\mathcal{N}=(0,4)$ supersymmetry.\footnote{Interesting proposals for the IR sigma model of the D1-D5 system probing the A-type ALE and ALF spaces were studied in \cite{Dabholkar:2008zy}.} In order to preserve supersymmetry in 2d it is necessary to perform a topological twist. Before proceeding with performing the twist, counting fields in the 2d theory and evaluating the central charges and levels, we first discuss the representations of the field content in the 4d theory.

\begin{table}[h]
\begin{centering}
\def\arraystretch{1.35}
\begin{tabular}{ccc}
\specialrule{.09em}{0.05em}{0em}
$\Gamma$ & Gauge multiplets & Hypermultiplets  \tabularnewline 
\specialrule{.05em}{0.08em}{0em}
$\mathbb{Z}_{m}$ & $U(1)^{m}$ & $m\times(\bs 1,\bs 1)$  \\
 
$\mathbb D_{m}^{*}$ & $U(1)^{4}\!\times \! U(2)^{m-1}$ & $4\times(\bs 1,\bs 2)+\left(m-2\right)\times(\bs 2,\bs 2)$  \\
 
$\mathbb T^{*}$ & $U(1)^{3}\!\times \!U(2)^{3}\! \times \! U(3)$ & $3\times(\bs 1,\bs 2)+3\times(\bs 2,\bs 3)$  \\
 
$\mathbb O^{*}$ & $U(1)^{2}\!\times\! U(2)^{3}\!\times \!U(3)^{2}\times \!U(4)$ & $2\times\big[(\bs 1,\bs 2)+(\bs 2,\bs 3)+(\bs 3,\bs 4)\big]+(\bs 2,\bs 4)$  \\
 
{$\mathbb I^{*}$} &{$U(1)\!\times\! U(2)^{2}\!\times\! U(3)^{2}\!\times\! U(4)^{2}\!\times\! U(5)\times\! U(6)$}  & {$(\bs 1,\bs 2)+(\bs 2,\bs 3)+(\bs 3,\bs 4)+(\bs 4,\bs 5)+$}  \tabularnewline
 &  & \!\!\!\!$(\bs 5,\bs 6)+(\bs 2,\bs 4)+(\bs 4,\bs 6)+(\bs 3,\bs 6)$   \\
\specialrule{.09em}{0.05em}{0em}
\end{tabular}
\par\end{centering}
\caption{\label{tab:Quivers-for-the}$\cN=2$ multiplets for the different quivers associated to the groups $\Gamma$.}
\end{table}

A 4d $\mathcal{N}=2$ SCFT admits an $SU(2)_{R}\times U(1)_{r}$ R-symmetry. The supercharges transform under the total symmetry group $SO(1,3)_\ell \times SU(2)_R\times U(1)_r$ in the representations
\begin{equation}
Q_{\ax I} \in (\bs 2, \bs 1, \bs 2)_{1}\, ,\qquad \tilde Q_{\dot \ax}^I\in (\bs 1, \bs 2, \bs 2)_{-1}\, .
\end{equation}
Let us consider a generic 4d $\mathcal{N}=2$ SCFT with $n_{H}$ $\mathcal{N}=2$ hypermultiplets and $n_V$ $\cN=2$ vector multiplets. An $\cN=2$ hypermultiplet has field content
\begin{equation}
q^I \in (\bs 1, \bs 1, \bs 2)_0\, , \quad 
\psi^{i=1,2}_\alpha \in (\bs 2, \bs 1, \bs 1)_{-1}\, ,\quad
\tilde \psi^{i=1,2}_{\dot \alpha} \in (\bs 1, \bs 2, \bs 1)_{1}\, ,
\end{equation}
whilst an $\cN=2$ vector multiplet contains the fields
\begin{equation}
A_\mu \in (\bs 2, \bs2, \bs 1)_0\, ,\quad 
\lambda^I_{\ax} \in (\bs 2, \bs 1, \bs 2)_1\, ,\quad 
\tilde \lambda^I_{\dot \ax} \in(\bs 1, \bs 2, \bs 2)_{-1}\, , \quad 
\Phi \in (\bs 1, \bs 1, \bs 1)_{2}\, .
\end{equation}
Note that in the case of the A-series we expect the 4d $\mathcal{N}=2$ SCFT to enjoy an additional $U(1)_L$ global symmetry. This flavor symmetry manifests itself in the D-brane picture as a $U(1)_L$ isometry of the transverse ALE space under which the supercharges are invariant, and is absent for the D- and E-series. The discussion of the assignment of $U(1)_L$ charges for the $\cN=2$ fields is relegated to appendix \ref{ULapp}. They will appear later in this section when we give the final 2d spectrum in table \ref{2dspectrum}.

Placing the 4d theories on $\mathbb{R}^{1,1}\times C$ breaks Lorentz symmetry with the group decomposing as $SO(1,3)_{\ell} \rightarrow SO(1,1)\times U(1)_{C}$. This splitting leads to the following decompositions of the $SO(1,3)_{\ell}$ representations:
\begin{align}
SO(1,3)_\ell \qquad &\rightarrow \qquad SO(1,1)\times U(1)_C\nn\\
(\bs 2,\bs 2)\qquad & \rightarrow \qquad \bs 1_{2,0}\oplus \bs 1_{-2, 0}\oplus \bs 1_{0,2}\oplus \bs 1_{0,-2}\nn\\
(\bs 2, \bs 1)\qquad  & \rightarrow \qquad \bs 1_{1,1}\oplus \bs 1_{-1, -1}\\
(\bs 1, \bs 2)\qquad  & \rightarrow \qquad \bs 1_{1,-1}\oplus \bs 1_{-1, 1}\, .\nn
\end{align}
The decomposition of the supercharges under $SU(2)_R \times SO(1,1)\times U(1)_C\times U(1)_r$ is:
\begin{align}
SO(1,3)_\ell \times SU(2)_R\times U(1)_r \qquad &\to \qquad SU(2)_R \times SO(1,1)\times U(1)_C\times U(1)_r\nn\\
Q_{\ax I}\in (\bs 2, \bs 1, \bs 2)_{1} \qquad &\to \qquad \bs 2_{1,1,1}\oplus \bs 2_{-1,-1,1}\\
\tilde Q_{\dot \ax}^I\in (\bs 1, \bs 2, \bs 2)_{-1} \qquad &\to \qquad \bs 2_{1,-1,-1}\oplus \bs 2_{-1, 1,-1}\, .\nn
\end{align}
In order to preserve $\cN=(0,4)$ in the 2d theory we must perform a topological twist with respect to $U(1)_r$ \cite{Bershadsky:1995vm,Kapustin:2006hi, Putrov:2015jpa, Amariti:2017cyd}. The unique twist preserving $\mathcal{N}=(0,4)$ in 2d, up to redefinition, is implemented by defining the generator of $U(1)'_C$ to be
\begin{equation}
T_C'=\frac{1}{2}\big(T_C+T_r \big)\, .
\end{equation}
In this way we obtain two singlet supercharges, both with negative chirality thereby giving $\cN=(0,4)$ supersymmetry in 2d. Next consider the decomposition of the various matter fields after the twist. The fields of an $\cN=2$ hypermultiplet decompose as:
\begin{align}\label{decomposition hypermultiplet}
SO(1,3)_\ell \times SU(2)_R\times U(1)_r \qquad &\to \qquad SU(2)_R \times SO(1,1)\times U(1)'_C\times U(1)_r\nn\\
q^I \in (\bs 1, \bs 1, \bs 2)_0 \qquad &\to \qquad  \bs 2_{0,0,0}\equiv q^I\nn\\
\psi^{i=1,2}_{\alpha} \in (\bs 2, \bs 1, \bs 1)_{-1} \qquad &\to \qquad \bs1_{1,0,-1}\oplus \bs 1_{-1,-1,-1}\equiv \psi^{i=1,2}_{+}\oplus \psi^{i=1,2} _{-}\\
\tilde \psi^{i=1,2}_{\dot \alpha} \in (\bs 1, \bs 2, \bs 1)_{1} \qquad &\to \qquad \bs 1_{1,0, 1}\oplus\bs 1_{-1, 1, 1}\equiv \tilde \psi^{i=1,2}_{+}\oplus \tilde \psi^{i=1,2 }_{-}\, .\nn
\end{align}
Likewise a 4d $\cN=2$ vector multiplet decomposes as:
\begin{align}\label{decomposition vector multiplet}
SO(1,3)_\ell \times SU(2)_R\times U(1)_r \qquad &\to \qquad SU(2)_R \times SO(1,1)\times U(1)'_C\times U(1)_r\nn\\
A_\mu \in (\bs 2, \bs2, \bs 1)_0 \qquad & \to \qquad \bs 1_{2,0,0}\oplus \bs 1_{-2, 0,0}\oplus \bs 1_{0,1,0}\oplus \bs 1_{0,-1,0}\nn\\
&\qquad \quad ~  \equiv v_+ \oplus v_-\oplus a \oplus \tilde a\nn\\
\lambda^I_{\ax} \in (\bs 2, \bs 1, \bs 2)_1\qquad &\to \qquad   \bs 2_{1,1,1}\oplus \bs 2_{-1,0,1}   \equiv \lambda^I_+\oplus \lambda^I_-     \\
\tilde \lambda^I_{\dot \ax} \in (\bs 1, \bs 2, \bs 2)_{-1} \qquad &\to \qquad \bs 2_{1,-1,-1}\oplus \bs 2_{-1,0,-1}\equiv \tilde \lambda^I_+ \oplus \tilde \lambda^I_-\nn\\
\Phi \in (\bs 1, \bs 1, \bs 1)_{2} \qquad &\to \qquad \bs 1_{0,1,2}\equiv \Phi\, .\nn
\end{align}

Having determined the 2d fields we now turn our attention to evaluating their multiplicities. 
The 2d fields with $U(1)'_C$ charges $Q'_C=\pm 1$ have multiplicities
\begin{equation}
\text{dim }H^0(C, K_C)=\text{dim }H^0(C, K^{-1}_C)=\text{dim }H^1(C, \cO_C)=g\, ,
\end{equation}
whereas the states with charge $Q'_C=0$ have multiplicity $\text{dim }H^0(C, \cO_C)=1$. The latter identity follows since the only holomorphic functions on a compact Riemann surface are constants. The full 2d massless spectrum follows from the decompositions \eqref{decomposition hypermultiplet} and \eqref{decomposition vector multiplet} coupled with the above discussion of the multiplicities, and is given in table \ref{2dspectrum}. We have included in table \ref{2dspectrum} the $U(1)_{L}$ flavor charge which is only present for the A-series case. We remind the reader that the details of the $U(1)_{L}$ charge assignment may be found in appendix \ref{ULapp}.
\begin{table}[h]
\begin{center}
\renewcommand{\arraystretch}{1.20}
\begin{tabular}[h]{ccc}
\specialrule{.09em}{0.05em}{0em}
Bosons & $SU(2)_R \times U(1)_L$ &    Multiplicity\\\specialrule{.07em}{0.05em}{0em}
$a, \tilde a, \Phi$  &   $2 \times \bs 1_0, \, 2 \times \bs 1_0$    &    $g$\\
$q^I$   &    $\bs 2_{1}, \bs2_{-1}$    &      1\\
$v_+, v_-$     &    $2\times \bs 1_0$ &    1\\
\specialrule{.09em}{0.05em}{0em}
Fermions & & \\\specialrule{.07em}{0.05em}{0em}
$\lambda^I_{+}, \tilde \lambda^I_+$   &   $2\times \bs 2_0$   &   $g$\\
$\psi^1_{+}, \psi^2_{+}, \tilde \psi^1_{+}, \tilde \psi^2_{+}$   &     $2 \times \bs 1_1, 2 \times \bs 1_{-1}$       &     1\\\hdashline
$\psi^1_{-}, \psi^2_{-}, \tilde \psi^1_{-}, \tilde \psi^2_{-}$   &     $2\times \bs 1_1, 2\times \bs 1_{-1}$       &     $g$\\
$\lambda^I_-,\tilde \lambda^I_-$   &      $2\times \bs 2_0$         &         $1$\\
\specialrule{.09em}{0.05em}{0em}
\end{tabular}
\end{center}
\caption{Spectrum of the 2d $\cN=(0,4)$ theory. The $U(1)_L$ charges are only relevant for the A-series.}
\lab{2dspectrum}
\end{table}

With the full 2d spectrum in hand we may compute 't Hooft anomaly coefficients and use their relation to the central charges and levels, see the recent paper \cite{Benini:2013cda} in the context of $\mathcal{N}=(0,2)$ theories as an example. The gravitational anomaly is easily computed via
\begin{equation}
c_L-c_R=\tr_{\text{Weyl}} \,\gamma_3=0\, ,
\end{equation}
where $\gamma_3$ is the chirality matrix in 2d with eigenvalues $\pm 1$ and the sum is over all Weyl fermions.

The right level is (recall $n_V=n_H=|\Gamma|$)
\begin{align}
k_R&=\tr_{\text{Weyl}} \big(\gamma_3 Q_{R}^2 \big)=|\Gamma|\bigg[2g \times \Big[\big(\tfrac{1}{2}\big)^2+\big(-\tfrac{1}{2}\big)^2 \Big]-2 \times \Big[\big(\tfrac{1}{2}\big)^2+\big(-\tfrac{1}{2}\big)^2 \Big] \bigg]\nn\\
&=|\Gamma|(g-1)=\frac{1}{2}|\Gamma| C \cdot C\, ,
\end{align}
where $Q_R$ is the charge under the Cartan of $SU(2)_R$. In the final line we have used adjunction which implies $g=\tfrac{1}{2}C\cdot C+1$ in order to express the level in terms of $C\cdot C$.

Proceeding in the same way to compute the 't Hooft anomaly associated to $U(1)_L$ we find\footnote{We compute the coefficient with half-integer $U(1)_L$ charges, with the embedding $U(1)_L \subset SU(2)_L$ in mind.}
\begin{align}
k_L=-\tr_{\text{Weyl}} \big(\gamma_3 Q_L^2 \big)=|\Gamma|(g-1)=\frac{1}{2}|\Gamma| C \cdot C\, .
\end{align}
The minus sign appearing in the definition of $k_{L}$, in contrast to the definition of $k_{R}$, arises as $U(1)_{L}$ is a left-moving current whilst $SU(2)_{R}$ is a right-moving current. Unitarity imposes that the anomalies are positive semi-definite, and the minus sign for the left-moving 't Hooft anomaly is included to `cancel off' the minus sign arising from the chirality matrix. Our 't Hooft anomaly considerations thus give
\begin{align}\label{microscopic central charges ale}
k_R=\frac{1}{2} |\Gamma| C \cdot C\, ,\qquad \qquad  k_L=\frac{1}{2}|\Gamma| C \cdot C\, ,\qquad\qquad  c_{L}-c_{R}=0\, ,
\end{align}
which precisely matches the result from the macroscopic computation \eqref{ALEmacroresults}.

We would now like to exploit the relation $c_R=6k_R=3 |\Gamma| C\cdot C$ dictated by supersymmetry to determine the central charges. However, this only holds if one can correctly identify the $SU(2)_R$ with the 2d R-symmetry and as already noted in section \ref{sec:6dstrings}, this does not work for the center of mass modes. This can be seen explicitly from table \ref{2dspectrum}. The center of mass modes are given by $q^I$, $\psi_{+}^{1,2}$, $\tilde{\psi}_+^{1,2}$, $\lambda^I_-$ and $\tilde{\lambda}^I_-$. For example, it is easy to see that $\lambda^I_-$ and $\tilde{\lambda}^I_-$ are left-moving but transform under $SU(2)_R$ which is forbidden if it is the (right-moving) R-symmetry. 
One should repeat these computations ignoring the center of mass modes in the traces. This yields the following:
\begin{equation}
c_L=c_R=6k_R=3|\Gamma|C\cdot C+6|\Gamma|\,,
\end{equation}
which are the central charges of the theory without the center of mass modes and is consistent with the results in \cite{ArabiArdehali:2018mil,Beccaria:2014qea} for $|\Gamma|=1$.

The central charges also follow from the relations $c_{L,R}=N_B^{L,R}+\tfrac{1}{2}N_F^{L,R}$, where $N_{B}^{L,R}$ is the number of left- and right-moving bosonic degrees of freedoms and $N_{F}^{L,R}$ the number of left- and right-moving fermionic degrees of freedom after extracting the center of mass degrees of freedom.

\subsection{Transverse ALF spaces: AD-series}

We now turn to studying the microscopics of the wrapped D3-brane probing a transverse ALF space. As there is strictly speaking no known worldvolume theory for this setting, unlike in the ALE case, we use a convenient duality frame to determine the central charges and levels. We first do this for the A-series where the ALF space is given by Taub-NUT space with NUT-charge $m$. Dualizing to an M5-brane picture allows us to compute the microscopic data. Unfortunately, for the D-series this method runs into difficulties which we shall explain in more detail below.

\paragraph*{A-series.}
To obtain an M-theory picture we first T-dualize along the NUT-circle, obtaining a type IIA setup with a D4-brane wrapping $C$ and the NUT-circle (with inverted radius), and $m$ NS5-branes wrapping the aforementioned circle and the entire K3. Uplifting to M-theory, we obtain M-theory compactified on $X=\text{K3}\times T^2$ with an M5-brane wrapping $C\times T^2$ and $m$ M5-branes wrapping K3. As explained in \cite{Bena:2006qm}, this system can effectively be described by an M5-brane wrapping $\mathcal P=C\times T^2+m \text{K3}$. This is the well-studied MSW CFT, where the Calabi--Yau threefold has been specialized to K3$\times T^2$. This leads to the added complication that there now exist one-cycles in the geometry which were assumed not to be present in the original MSW setup, as the Calabi--Yau was taken to have $SU(3)$ holonomy. In order to determine the microscopic central charges we compute the bosonic and fermionic spectrum in the effective 2d $\cN=(0,4)$ theory on the M5-brane. This setup was already studied in  \cite{LopesCardoso:1999fsj,Lambert:2007is,Dabholkar:2010rm} before, but for convenience and completeness we review the main points here.

The worldvolume theory of the M5-brane is a 6d $(2,0)$ theory with a $(2,0)$ tensor multiplet containing a chiral two-form and five scalar fields. Three of the five scalars parametrize the center of mass motion of the string in the transverse $\mathbb{R}^3$, whereas the remaining two scalars parametrize the position of the curve $\mathcal P$ inside $X=\text{K3}\times T^2$. Additionally, one obtains 2d scalar fields from reducing the chiral two-form along two-forms on $\mathcal P$. Due to the self-duality constraint there will be $b_2^-(\mathcal P)$ left-moving, and $b_2^+(\mathcal P)$ right-moving scalar fields in two dimensions, where $b_2^\pm(\mathcal P)$ denote the number of harmonic (anti-)self-dual two-forms on $\mathcal P$. In total, we find
\begin{align}
N^L_{B}=d_{\mathcal P}+3+b_2^-(\mathcal P)\, ,\qquad N^R_{B}=d_{\mathcal P}+3+b_2^+(\mathcal P)
\end{align}
left- and right-moving bosons. The contribution from the two scalars parametrizing the motion of $\mathcal P$ inside K3$\times T^2$ is denoted by $d_{\mathcal P}$ and will be determined momentarily. The number of left- and right-moving fermions can be determined in the standard way by using the formulas
\begin{align}
N^L_{F}&=4 h^{0,1}(\mathcal P)=2 b_1(\mathcal P)=2 b_1(X)=4\, ,\\
N^R_{F}&=4(b_0(\mathcal P)+h^{0,2}(\mathcal P))=4\big(b_0(X)+h^{0,2}(\mathcal P)\big)=4+4 h^{0,2}(\mathcal P)\, .\nn
\end{align} 
Supersymmetry in the right-moving sector implies that the number of right-moving bosons equals the number of right-moving fermions, allowing us to determine $d_{\mathcal P}$ via
\begin{align}
0=N^R_{B}-N^R_{F}=d_{\mathcal P}+b_2^+(\mathcal P)-4h^{0,2}(\mathcal P)-1=d_{\mathcal P}-2h^{0,2}(\mathcal P)\, ,
\end{align}
where we have used that $b_2^+(\mathcal P)=b_2(\mathcal P)-b_2^-(\mathcal P)$ and $b_2^-(\mathcal P)=h^{1,1}(\mathcal P)-1$. For the left- and right-moving central charges we find,
\begin{equation}
c_L=b_2(\mathcal P)+4 \, ,\qquad c_R=3b_2^+(\mathcal P)+3\, .
\end{equation}
We may evaluate these explicitly by rewriting the topological numbers in terms of integrals on $X=\text{K3}\times T^2$ as \cite{Maldacena:1997de,Dabholkar:2010rm}
\begin{equation}
c_L=\int_{X}\big({\mathcal P}^3+c_2(X)\wedge \mathcal P\big)+6\, ,\qquad 
c_R=\int_{X}\Big({\mathcal P}^3+\frac{1}{2}c_2(X)\wedge\mathcal P\Big)+6\, .
\end{equation}
These are straightforward to evaluate for the curve $\mathcal{P}$, giving the results\footnote{When one calculates the central charges using 't Hooft anomalies and the relation $c_R=6k_R$ one finds the results \eqref{Aseriesmicroresults} up to order $\mathcal{O}(1)$ \cite{Dabholkar:2010rm}. The difference is again caused by subtleties in identifying $SU(2)_R$ with the R-symmetry.}:
\begin{align}
c_L=3m C\cdot C+24m+6\, ,\qquad c_R=3m C\cdot C+12m+6\, .\lab{Aseriesmicroresults}
\end{align}
Using $c_R=6k_R$ this matches the macroscopic result \eqref{Aseriesmacroresults} up to and including linear order terms in the charges. We expect that the left level can be computed by adapting the techniques in \cite{Bena:2006qm,Grimm:2018weo}.

\paragraph*{D-series.}

We saw above that the microscopics in the A-series' case is relatively clean once we have dualized to an M-theory picture. The D-series' case on the other hand, is not as clean cut. This is because the dual M-theory setup is no longer accessible with the standard MSW techniques. To see why, let us perform the analogous duality chain as in the A-series. Performing a T-duality along the fiber of the D-series ALF space to type IIA we obtain a D4-brane wrapping S$^1\times C\times$S$_{\mathrm{D}}^1$ from the D3-brane, $m$ NS5-branes wrapping K3$\times$S$^1$ and in addition a so called ON$5$-plane on top of the NS5-branes \cite{Hanany:2000fq}. An ON$5$-plane is the analogue of an ordinary orientifold plane for the NS5-brane, and leads to orthogonal gauge groups when placed on top of the NS5-branes. Lifting the type IIA setup to M-theory the D4 lifts to an M5-brane wrapping $C\times T^2$, the NS5s to M5s on K3 whilst the ON$5$-plane becomes an OM$5$ plane wrapping the K3. At low energies the system of M5-branes with an OM$5$-plane on top realize the familiar D-type $(2,0)$ SCFT in the ADE classification. This is a very involved setup and does not directly lead itself to the application of MSW to compute the central charges due to the presence of the orientifold.

A second chain of dualities, which can also be applied to all the other cases considered in this paper, is to first T-dualize along the circle wrapped by the D3-brane, leading to a type IIA setup of a D2-brane probing the ALE/ALF space. Uplifting to M-theory we obtain an M2-brane on the curve $C$ probing the ALE/ALF space. The problem has now been rephrased in terms of M2-brane counting for five-dimensional black holes. Note that as this duality frame exists for all the cases considered here, it is to some extent the most universal setup. In flat space there is a connection between the black hole partition function and the topological string partition function. It is natural to conjecture that a generalization of this connection would be a useful tool in understanding the counting of M2-branes in this setup.

Since the center of the ALF space looks like $\mathbb{C}^2/\mathbb{D}_m^*$ it is not unreasonable to expect that the leading order contribution to the central charges and level agrees with the corresponding D-series ALE space. The macroscopic results \eqref{Dmacro} confirm this expectation. Clearly it is desirable to have a better understanding of this case and to obtain a first principles derivation of the microscopic central charges and level.


\section{Conclusion and outlook}\label{sec:conclusions}
We studied the central charges and levels of 2d $\mathcal{N}=(0,4)$ SCFTs dual to black strings that arise from D3-branes in type IIB compactifications on K3. The branes wrap S$^1\times C$, where $C$ is a curve in K3, and have as transverse space an ALE or ALF space. We computed the central charges and levels both from six-dimensional $\mathcal{N}=(2,0)$ supergravity as from the $\mathcal{N}=(0,4)$ SCFTs. We found excellent matching between microscopics and macroscopics for all ALE spaces and for ALF spaces corresponding to the A-series. For ALF spaces corresponding to the D-series we only performed a macroscopic analysis.

A natural extension of this work is to in addition include 7-branes in the setup. The natural framework for this is F-theory \cite{Vafa:1996xn}. Whereas in this paper we have compactified type IIB on K3, wrapping the D3-brane on a curve $C$ inside K3, we would instead compactify F-theory on an elliptically fibered Calabi--Yau threefold, wrapping the D3-brane on a curve $C$ inside the base of the threefold. This has been considered for flat transverse space in \cite{Haghighat:2015ega,Lawrie:2016axq} and for Taub-NUT in \cite{Bena:2006qm, Couzens:2017way, Grimm:2018weo}. The macroscopic computations for Taub-NUT performed in \cite{Grimm:2018weo} straightforwardly extend to the ALE and ALF transverse spaces considered here. Microscopically this requires a modification of the topological duality twist \cite{Martucci:2014ema,Assel:2016wcr,Lawrie:2016axq} which is an interesting problem that we hope to turn our attention to in the future.

\paragraph*{Acknowledgements.}

It is a pleasure to thank
Markus Dierigl, Nava Gaddam, Thomas Grimm, Miguel Montero, and Yuji Tachikawa
for valuable discussions and correspondence. KM would like to thank Kavli IPMU Tokyo for their kind hospitality whilst part of this work was completed.

This work was supported in part by the D-ITP consortium, a program of the Netherlands Organization for Scientific Research (NWO) that is funded by the Dutch Ministry of Education,
Culture and Science (OCW), and by the NWO Graduate Programme.

\newpage

\appendix


\section{Action of $\mathbb{Z}_{m}$ and $\mathbb{D}_{m}^{*}$ on KK spectrum \label{sec:Action An and Dn on KK spectrum}}

In order to compute the one-loop quantum corrections for the ALF transverse spaces we must determine the spectrum on AdS$_3\times$S$^{3}/\Gamma$. This is achieved by taking the spectrum of AdS$_3\times$S$^3$, as given in \cite{Deger:1998nm,deBoer:1998kjm}, and truncating out modes that are \emph{not} invariant under $\Gamma \subset SU(2)_L$. We shall consider $\Gamma=\mathbb{Z}_{m}$ and $\Gamma=\mathbb{D}^{*}_{m}$ in turn. The representations are determined by $SU(2)_{L}$ representations labeled by $j_{L}$ with $j_{L}\in \frac{1}{2} \mathbb{N}$ and with dimension $2 j_{L}+1$. The $j_{L}$ representation is given by the $2 j_{L}$-fold symmetrized tensor product of the fundamental representation ($j_{L}=\frac{1}{2}$). 
The fundamental representation is given by the standard action of $SU(2)_L$ on $\mathbb{C}^2$. Let us denote the basis of $\mathbb{C}^2$ by
\begin{equation}
\ket{\uparrow}=
\begin{pmatrix}
1\\
0
\end{pmatrix}
\, , \qquad \ket{\downarrow}=
\begin{pmatrix}
0\\
1
\end{pmatrix}\, .
\end{equation}
The vector space $V_{j_L}$ corresponding to the representation labeled by $j_L$ is given by $V_{j_L}=\text{Sym}^{2j_L}(\mathbb{C}^2)$. A basis for this vector space is given by 
\begin{equation}
\ket{j_L, j_L^3}=\text{Sym}^{2j_L}(\uparrow^{j_L+j_L^{3}} \, \downarrow^{j_L-j_L^{3}})\equiv \underbrace{\ket{\uparrow} \otimes \cdots \otimes \ket{\uparrow}}_{(j_L+j_L^{3}) \text{ factors} } \otimes \underbrace{ \ket{\downarrow} \otimes \cdots \otimes \ket{\downarrow}}_{(j_L-j_L^{3}) \text{ factors}}\quad  + \quad  \text{symmetrized}\, ,
\end{equation}
for $j_L^3=-j_L, \,-j_L+1, \dots, j_L$. Consider now the actions of $\mathbb{Z}_{m}$ and $\mathbb{D}^{*}_{m}$ on the above representations.

\paragraph*{Action $\mathbb{Z}_m$.}
The action of the group $\mathbb{Z}_{m}\subset SU(2)_L$ on the fundamental representation
is generated by 
\begin{equation}
\mathcal{A}=\left(\begin{array}{cc}
\e^{\frac{2i\pi}{m}} & 0\\
0 & \e^{-\frac{2i\pi}{m}}
\end{array}\right).
\end{equation}
Using this action and the construction of the other representations in terms of the fundamental one, we can explicitly deduce how the basis states $\ket{j_L, j_L^3}$ transform under the action of $\mathcal A$. We find 
\begin{equation}
 ~ \ket{j_L, j_L^3} \overset{\mathcal{A}}{\longrightarrow} ~ \e^{4\pi i \frac{j_L^{3}}{m}} \ket{j_L, j_L^3}\, .
\end{equation}
The modes invariant under $\mathbb{Z}_m$ are those with $j_L^{3}=\frac{1}{2}m k$ for $k \in \mathbb{Z}$. 

\paragraph*{Action $\mathbb{D}_m^*$.}

Consider now the action of $\mathbb{D}_m^*\subset SU(2)_L$ on the fundamental representation. It is generated by the two generators
\begin{equation}
\mathcal{A}=\left(\begin{array}{cc}
\e^{\frac{i\pi}{m}} & 0\\
0 & \e^{-\frac{i\pi}{m}}
\end{array}\right)\,,\qquad\mathcal{B}=\left(\begin{array}{cc}
0 & i\\
i & 0
\end{array}\right)\,.\label{gens}
\end{equation}
Similarly we find that under the action of the generator $\mathcal{A}$
\begin{equation}
 ~ \ket{j_L, j_L^3} \overset{\mathcal{A}}{\longrightarrow} ~ \e^{2\pi i \frac{j_L^{3}}{m}} \ket{j_L, j_L^3}\, ,
\end{equation}
implying the projection condition $j_L^{3}=m k$ for $k \in \mathbb{Z}$. Note that this also implies $j_L \in \mathbb{N}$. For the generator $\mathcal{B}$ we find
\begin{equation}
\ket{j_L, j_L^3}~ \overset{\mathcal{B}}{\longrightarrow}~(-1)^{j_L} \ket{j ,-j_L^3}\, .
\end{equation}
To obtain invariant states under the generator $\mathcal{B}$ we construct the linear combinations
\begin{equation}
\ket{j_L, j_L^{3}}+(-1)^{j_L} \ket{{j_L, -j_L^{3}}} \, .
\end{equation}
These states are then invariant under both generators provided $j_{L}^{3}=mk$ for $k\in \mathbb{Z}_{\geq 0}$.


\section{Summation of 3d one-loop corrections \label{sec:Summation-of-3D one-loop corrections}}

In this appendix we present the full and explicit computation of the one-loop contributions to the three-dimensional $\mathfrak{u}(1)_{L},$ $\mathfrak{su}(2)_{R}$ and gravitational Chern--Simons terms from integrating out the massive KK modes arising in the reduction on S$^3/\mathbb{Z}_{m}$. With these results, and a little further work, we may also obtain the results for the reduction on S$^{3}/\mathbb{D}^{*}_{m}$. For this reason in this appendix we will almost exclusively consider the A-series case. The necessary adaptation to the D-series case will be explained in the final section \ref{D-series' corrections}. 

To determine the corrections furnished by the massive KK modes we sum the contributions of individual modes (see table \ref{looptable}) over the full KK spectrum derived in section \ref{subsec:Kaluza-Klein-spectrum}. As explained in the main text we must regularize the summations using zeta-function regularization. In particular we will use the following regularized summations:
\begin{eqnarray}
\sum_{j_{L}=\frac{1}{2}mk}^{\infty}1 & = & \frac{1}{2}-\frac{1}{2}mk\,,\qquad\sum_{j_{L}=\frac{1}{2}mk}^{\infty}j_{L}=\frac{1}{24}\big(-2+6km-3k^{2}m^{2}\big)\,,\nonumber \\
\sum_{j_{L}=\frac{1}{2}mk}^{\infty}j_{L}^{2} & = & \frac{1}{24}\big(-2km+3k^{2}m^{2}-k^{3}m^{3}\big)\,.
\end{eqnarray}
The sums are performed over the integers or half integers when $\tfrac{1}{2}mk$ is integer or half integer respectively. In fact the regularized summations we require are:\begin{eqnarray}
\sum_{k=1}^{\infty}\sum_{j_{L}=\frac{1}{2}mk}^{\infty}1 & = & -\frac{1}{4}+\frac{m}{24}\,,\qquad\qquad\quad\;\sum_{k=1}^{\infty}\sum_{j_{L}=\frac{1}{2}mk}^{\infty}j_{L}=\frac{1}{24}-\frac{m}{48}\,,\nn \\
\sum_{k=1}^{\infty}\sum_{j_{L}=\frac{1}{2}mk}^{\infty}j_{L}^{2} & = & -\frac{m^{3}}{24\cdot120}+\frac{m}{144}\,,\qquad\sum_{k=1}^{\infty}\sum_{j_{L}=\frac{1}{2}mk}^{\infty}k^{2}=-\frac{m}{240}\,,
\end{eqnarray}
where we have used the previous results and (\ref{zeta-function regularization}).

We now calculate the corrections $k_{L}^{\mathrm{loop}},$ $k_{R}^{\mathrm{loop}}$
and $(c_{L}-c_{R})^{\mathrm{loop}}$ separately. As described in the
main text we implement the projection condition $j_{L}=\tfrac{1}{2}mk$
for $k\in\mathbb{Z}_{\geq0}$ by first summing over the representations
$j_{L}=\tfrac{1}{2}mk,$ $\tfrac{1}{2}mk+1,$ $...$ and subsequently
summing over $k$. The structure of the KK spectrum is such that we
have to restrict to $m\geq3$ and do the sums for $k=0$ separately.
In the calculation of the one-loop contribution to $k_L$, $k_R$ and $c_L-c_R$ we therefore first sum over the representations relevant for $k=0$ and afterwards we sum the contributions for $k\neq0.$ We will not perform the computation for the special cases $m=1,$ $2$ explicitly as they give the same result up to order $\mathcal{O}(1)$ as the formulas for $m\geq 3$ evaluated in $m=1,2$.

\subsection{Relevant spectrum}

The part of the spectrum that contributes to $k=0$ is given by 
\begin{itemize}
\item Spin-$\frac{3}{2}$:
\[
4\bigoplus_{j_{L}=1}^{\infty}\big(j_{L},j_{L}\pm\tfrac{1}{2}\big)^{\mp}.
\]
\item Spin-$\frac{1}{2}$:
\begin{eqnarray*}
4\bigoplus_{j_{L}=2}^{\infty}\big(j_{L},j_{L}\pm\tfrac{3}{2}\big)^{\mp} & \oplus & 4\bigoplus_{j_{L}=0}^{1}\big(j_{L},j_{L}+\tfrac{3}{2}\big)^{-}\oplus4\bigoplus_{j_{L}=1}^{\infty}\big(j_{L},j_{L}\pm\tfrac{1}{2}\big)^{\pm}\\
\oplus \, 4\big(0,\tfrac{1}{2}\big)^{+} & \oplus & 84 \bigoplus_{j_{L}=1}^{\infty}\big(j_{L},j_{L}\pm\tfrac{1}{2}\big)^{\pm}\oplus 84 \big(0,\tfrac{1}{2}\big)^{+}.
\end{eqnarray*}
\item Chiral vectors:
\[
5\bigoplus_{j_{L}=1}^{\infty}\left(j_{L},j_{L}\pm1\right)^{\mp}\oplus 5\left(0,1\right)^{-}\oplus 21 \bigoplus_{j_{L}=1}^{\infty}\left(j_{L},j_{L}\pm1\right)^{\pm}\oplus 21 \left(0,1\right)^{+}.
\]
\end{itemize}
where all the sums are over the integers.

The vector representations $(0,1)^{-}\oplus 21 (0,1)^{+}$ are mapped
to themselves by the reality condition on the six-dimensional tensors.
We thus count their contribution with a factor of $\tfrac{1}{2}.$ 

When $k>0,$ we find that $j_{L}^{3}=\frac{1}{2}mk\geq\tfrac{m}{2}$
such that for $m\geq3$ the following part of the spectrum contributes:
\begin{itemize}
\item Spin-$\frac{3}{2}$:
\[
4\bigoplus_{j_{L}=\frac{1}{2}mk}^{\infty}\big(j_{L},j_{L}\pm\tfrac{1}{2}\big)^{\mp}.
\]
\item Spin-$\frac{1}{2}$:
\[
4\bigoplus_{j_{L}=\frac{1}{2}mk}^{\infty}\big(j_{L},j_{L}\pm\tfrac{3}{2}\big)^{\mp}\oplus4\bigoplus_{j_{L}=\frac{1}{2}mk}^{\infty}\big(j_{L},j_{L}\pm\tfrac{1}{2}\big)^{\pm}\oplus 84 \bigoplus_{j_{L}=\frac{1}{2}mk}^{\infty}\big(j_{L},j_{L}\pm\tfrac{1}{2}\big)^{\pm}.
\]
\item Chiral vectors:
\[
5\bigoplus_{j_{L}=\frac{1}{2}mk}^{\infty}\left(j_{L},j_{L}\pm1\right)^{\mp}\oplus 21 \bigoplus_{j_{L}=\frac{1}{2}mk}^{\infty}\left(j_{L},j_{L}\pm1\right)^{\pm}.
\]
\end{itemize}
The sums are again with integer steps.

\subsection{Correction to the left level}

Note that the contribution from the $k=0$ modes is zero in this case as they come with an overall factor of $(j_L^3)^2$ which clearly vanishes in the $k=0$ case. We next consider, separately at first, the contributions for $k>0$ for the different
kinds of fields before summing all the results. For the spin-$\tfrac{3}{2}$ fermions we find
\begin{eqnarray}
\alpha_{L}^{\mathrm{(3/2)}} & = & 4\sum_{k=1}^{\infty}\sum_{j_{L}=\frac{1}{2}mk}^{\infty}\frac{3}{8\pi}\big(\tfrac{1}{2}mk\big)^{2}\big[2\big(j_{L}-\tfrac{1}{2}\big)+1-2\big(j_{L}+\tfrac{1}{2}\big)-1\big]\nonumber \\
 & = & -\frac{3m^{2}}{4\pi}\sum_{k=1}^{\infty}\sum_{j_{L}=\frac{1}{2}mk}^{\infty}k{}^{2}=\frac{1}{8\pi}\frac{m^{3}}{40}\,.\label{eq:left level contr spin 3/2}
\end{eqnarray}
Likewise the spin-$\tfrac{1}{2}$ fermions contribute as
\begin{eqnarray}
\alpha_{L}^{\mathrm{(1/2)}} & = & 4\sum_{k=1}^{\infty}\sum_{j_{L}=\frac{1}{2}mk}^{\infty}\frac{1}{8\pi}\big(\tfrac{1}{2}mk\big)^{2}(-6+2+42)\nonumber \\
 & = & \frac{19}{4\pi}m^{2}\sum_{k=1}^{\infty}\sum_{j_{L}=\frac{1}{2}mk}^{\infty}k^{2}=-\frac{1}{8\pi}\frac{19}{120}m^{3}\,.\label{left level contr spin 1/2}
\end{eqnarray}
Finally, the vectors contribute as
\begin{equation}
\alpha_{L}^{\mathrm{(vect)}}=\sum_{k=1}^{\infty}\sum_{j_{L}=\frac{1}{2}mk}^{\infty}\frac{1}{4\pi}\big(\tfrac{1}{2}mk\big)^{2}(20-84)=\frac{1}{8\pi}\frac{2}{15}m^{3}\,.\label{left level contr vectors}
\end{equation}
Adding (\ref{eq:left level contr spin 3/2}), (\ref{left level contr spin 1/2})
and (\ref{left level contr vectors}), it follows that the one-loop contribution to the left level vanishes:
\begin{eqnarray}
k_{L}^{\mathrm{loop}} & = & 8\pi\cdot\Big(\alpha_{L}^{\mathrm{(3/2)}}+\alpha_{L}^{\mathrm{(1/2)}}+\alpha_{L}^{\mathrm{(vect)}}\Big)\nonumber \\
 & = & 0\,.
\end{eqnarray}

\subsection{Correction to the right level}
Let us now perform the analogous computation for the right level. We begin by studying the contribution from the $k=0$ modes which do not vanish in this case. The spin-$\tfrac{3}{2}$
fermions contribute as
\begin{equation}
4\sum_{j_{L}=1}^{\infty}\frac{1}{4\pi}\big[\big(j_{L}+\tfrac{1}{2}\big)\big(j_{L}+\tfrac{3}{2}\big)\big(2j_{L}+2\big)-\big(j_{L}-\tfrac{1}{2}\big)\big(j_{L}+\tfrac{1}{2}\big)\big(2j_{L}\big)\big]=-\frac{5}{4\pi}\,.
\end{equation}
The infinite towers contained within the spin-$\tfrac{1}{2}$
fermion spectrum give
\begin{eqnarray}
\frac{1}{3\pi}\sum_{j_{L}=2}^{\infty}\big[\big(j_{L}+\tfrac{3}{2}\big)\big(j_{L}+\tfrac{5}{2})\big(2j_{L}+4\big)-\big(j_{L}-\tfrac{3}{2}\big)\big(j_{L}-\tfrac{1}{2})\big(2j_{L}-2\big)\big] & = & -\frac{83}{4\pi}\,,\nonumber \\
\frac{-1}{3\pi}(1+21)\sum_{j_{L}=1}^{\infty}\big[\big(j_{L}+\tfrac{1}{2}\big)\big(j_{L}+\tfrac{3}{2})\big(2j_{L}+2\big)-\big(j_{L}-\tfrac{1}{2}\big)\big(j_{L}+\tfrac{1}{2})\big(2j_{L}\big)\big] & = & \frac{55}{6\pi}\, ,\nonumber \\
\end{eqnarray}
whilst the isolated representations give
\begin{equation}
-\frac{1}{3\pi}\big[-\tfrac{3}{2}\cdot\tfrac{5}{2}\cdot4-\tfrac{5}{2}\cdot\tfrac{7}{2}\cdot6+(1+21)\cdot\tfrac{1}{2}\cdot\tfrac{3}{2}\cdot2\big]=\frac{23}{2\pi}\,.
\end{equation}
Finally, the infinite towers contained within the vector spectrum contribute
\begin{equation}
(5-21)\sum_{j_{L}=1}^{\infty}\frac{1}{6\pi}\big[-\big(j_{L}+1\big)\big(j_{L}+2)\big(2j_{L}+3\big)+\big(j_{L}-1\big)j_{L}\big(2j_{L}-1\big)\big]=-\frac{32}{3\pi}
\end{equation}
and the isolated representations $5\big(0,1\big)^{-}\oplus 21 \big(0,1\big)^{-}$
give (note that they come with an extra factor of $\tfrac{1}{2}$)
\begin{equation}
-\frac{1}{2}\left(5-21\right)\cdot \frac{1}{6\pi}\cdot1\cdot2\cdot3=\frac{8}{\pi}\,.
\end{equation}
Enumerating all the contributions  from the above $k=0$ results gives
\begin{equation}
\alpha_{R}^{k=0}=-\frac{16}{4\pi}\,.\label{right level k=00003D0}
\end{equation}

We now turn to the evaluation of the $k>0$ contributions. For the spin-$\tfrac{3}{2}$
fermions the contribution is
\begin{eqnarray}
\alpha_{R}^{\mathrm{(3/2)}} & = & 4\sum_{k=1}^{\infty}\sum_{j_{L}=\frac{1}{2}mk}^{\infty}\frac{1}{4\pi}\big[\big(j_{L}+\tfrac{1}{2}\big)\big(j_{L}+\tfrac{3}{2}\big)\big(2j_{L}+2\big)-\big(j_{L}-\tfrac{1}{2}\big)\big(j_{L}+\tfrac{1}{2}\big)\big(2j_{L}\big)\big]\nonumber \\
 & = & \frac{1}{\pi}\sum_{k=1}^{\infty}\sum_{j_{L}=\frac{1}{2}mk}^{\infty}\Big(\frac{3}{2}+6j_{L}+6j_{L}^{2}\Big)=\frac{1}{4\pi}\Big(-\frac{1}{2}-\frac{m}{12}-\frac{m^{3}}{120}\Big)\,.\label{right level 3/2 contribution}
\end{eqnarray}
Correspondingly, the spin-$\tfrac{1}{2}$ fermions give
\begin{eqnarray}
\alpha_{R}^{\mathrm{(1/2)}} & = & 4\sum_{k=1}^{\infty}\sum_{j_{L}=\frac{1}{2}mk}^{\infty}\frac{1}{12\pi}\big[\big(j_{L}+\tfrac{3}{2}\big)\big(j_{L}+\tfrac{5}{2})\big(2j_{L}+4\big)-\big(j_{L}-\tfrac{3}{2}\big)\big(j_{L}-\tfrac{1}{2})\big(2j_{L}-2\big)\big]\nonumber \\
 &  & -4(1+21)\sum_{k=1}^{\infty}\sum_{j_{L}=\frac{1}{2}mk}^{\infty}\frac{1}{12\pi}\big[\big(j_{L}+\tfrac{1}{2}\big)\big(j_{L}+\tfrac{3}{2})\big(2j_{L}+2\big)-\big(j_{L}-\tfrac{1}{2}\big)\big(j_{L}+\tfrac{1}{2})\big(2j_{L}\big)\big]\nonumber \\
 & = & \frac{1}{4\pi}\sum_{k=1}^{\infty}\sum_{j_{L}=\frac{1}{2}mk}^{\infty}\big(22+24j_{L}+24j_{L}^{2}\big)-\frac{22}{3\pi}\sum_{k=1}^{\infty}\sum_{j_{L}=\frac{1}{2}mk}^{\infty}\Big(\frac{3}{2}+6j_{L}+6j_{L}^{2}\Big)\nonumber \\
 & = & \frac{1}{4\pi}\Big(-\frac{5}{6}+\frac{43}{36}m+\frac{19}{360}m^{3}\Big)\,.\label{right level 1/2 contribution}
\end{eqnarray}
Lastly the vector spectrum provides the contribution 
\begin{eqnarray}
\alpha_{R}^{\mathrm{(vect)}} & = & (21-5)\sum_{k=1}^{\infty}\sum_{j_{L}=\frac{1}{2}mk}^{\infty}\frac{1}{6\pi}\big[\big(j_{L}+1\big)\big(j_{L}+2)\big(2j_{L}+3\big)-\big(j_{L}-1\big)j_{L}\big(2j_{L}-1\big)\big]\nonumber \\
 & = & -\frac{16}{\pi}\sum_{k=1}^{\infty}\sum_{j_{L}=\frac{1}{2}mk}^{\infty}\big(1+2j_{L}+2j_{L}^{2}\big)=\frac{1}{4\pi}\Big(\frac{32}{3}-\frac{8}{9}m+\frac{2}{45}m^{3}\Big)\,. \nn \\
 && \, \label{right level vect contribution}
\end{eqnarray}
Adding (\ref{right level k=00003D0}), (\ref{right level 3/2 contribution}),
(\ref{right level 1/2 contribution}) and (\ref{right level vect contribution})
the one-loop contribution to the right level is
\begin{eqnarray}
k_{R}^{\mathrm{loop}} & = & 4\pi\cdot\Big(\alpha_{R}^{k=0}+\alpha_{R}^{\mathrm{(3/2)}}+\alpha_{R}^{\mathrm{(1/2)}}+\alpha_{R}^{\mathrm{(vect)}}\Big)\nonumber \\
 & = & 2m-28\,.
\end{eqnarray}


\subsection{Correction to $c_{L}-c_{R}$}

As in the previous section we first calculate the $k=0$ contribution before proceeding to calculate the $k>0$ contributions. The spin-$\tfrac{3}{2}$ fermion, spin-$\tfrac{1}{2}$ fermion and vector representations give
\begin{eqnarray}
4\sum_{j_{L}=1}^{\infty}\frac{7}{64\pi}\cdot2 & = & -\frac{7}{16\pi}\,,\nonumber \\
\frac{1}{48\pi}\bigg[-\sum_{j_{L}=2}^{\infty}6-4-6+\sum_{j_{L}=1}^{\infty}2+2+21\sum_{j_{L}=1}^{\infty}2+42\bigg] & = & \frac{7}{16\pi}\,,\\
(21-5)\sum_{j_{L}=1}^{\infty}\frac{1}{48\pi}\cdot4+\frac{1}{2}(21-5)\frac{1}{48\pi}\cdot3 & = & \frac{1}{6\pi}\,,\nonumber 
\end{eqnarray}
respectively. We have again included an extra factor
of $\tfrac{1}{2}$ for the isolated vector representations. The $k=0$ contribution from the above corrections is
\begin{equation}
\alpha_{\mathrm{grav}}^{k=0}=-\frac{16}{96\pi}\,.\label{gravitational k=00003D0 contr}
\end{equation}
The $k>0$ contribution of the spin-$\tfrac{3}{2}$ fermions equals
\begin{equation}
\alpha_{\mathrm{grav}}^{\mathrm{(3/2)}}=4\sum_{k=1}^{\infty}\sum_{j_{L}=\frac{1}{2}mk}^{\infty}\frac{7}{64\pi}\cdot2=\frac{1}{96\pi}\Big(-21+\frac{7}{2}m\Big)\,,\label{grav spin 3/2 contr}
\end{equation}
the spin-$\tfrac{1}{2}$ fermions contribute
\begin{equation}
\alpha_{\mathrm{grav}}^{\mathrm{(1/2)}}=4\sum_{k=1}^{\infty}\sum_{j_{L}=\frac{1}{2}mk}^{\infty}\frac{1}{48\cdot4\pi}(-6+2+42)=\frac{1}{96\pi}\Big(-19+\frac{19}{6}m\Big)\,,\label{grav spin 1/2 contr}
\end{equation}
and finally the vector modes give
\begin{equation}
\alpha_{\mathrm{grav}}^{\mathrm{(vect)}}=(19-5)\sum_{k=1}^{\infty}\sum_{j_{L}=\frac{1}{2}mk}^{\infty}\frac{1}{48\pi}\cdot4=\frac{1}{96\pi}\Big(32-\frac{16}{3}m\Big)\,.\label{grav vect contr}
\end{equation}
The total one-loop correction for $c_{L}-c_{R}$, obtained by summing (\ref{gravitational k=00003D0 contr}), (\ref{grav spin 3/2 contr}), (\ref{grav spin 1/2 contr}) and (\ref{grav vect contr}) is
\begin{eqnarray}
(c_{L}-c_{R})^{\mathrm{loop}} & = & 96\pi\cdot\Big(\alpha_{\mathrm{grav}}^{k=0}+\alpha_{\mathrm{grav}}^{\mathrm{(3/2)}}+\alpha_{\mathrm{grav}}^{\mathrm{(1/2)}}+\alpha_{\mathrm{grav}}^{\mathrm{(vect)}}\Big)\nonumber \\
 & = & 12m-88\,.
\end{eqnarray}

\subsection{$\mathbb{D}^{*}_{m}$ quantum corrections from $\mathbb{Z}_{m}$} \label{D-series' corrections}

In the previous sections we have computed the one-loop corrections to the levels and gravitational central charge for the A-series. We must also compute the right level and gravitational central charge for the D-series. In principle one should be able to perform the equivalent computations of the previous sections, however in practice this is not the most efficient way to obtain the desired result. Instead we can use the observation that the summation in the D-series is equivalent, modulo the $j^{3}_{R}\geq 0$ condition, to the summations of the A-series with $m\rightarrow 2m$. We may therefore use the previous results and then impose the $j^{3}_{R}\geq 0$ condition. As the contribution for a given $j^{3}_{R}$ within a  fixed $j_{R}$ representation is the same for $j_{R}^{3}$ and $-j_{R}^{3}$ imposing $j_{R}^{3}\geq0$ can be performed by halving the above result and adding in $\tfrac{1}{2}$ of the contribution from $j_{R}^{3}=0$. The latter contributions can be seen to be of order 1 in the charges and therefore, as we are working to linear order, we may neglect this discrepancy and use the above shortcut. Performing the computation as outlined in the A-series section above for the D-series will result in the same linear order results. 
The final quantum corrections for the D-series are then
\begin{equation}
k_{R}^{\mathrm{loop}}=2m\, ,~~~~~(c_{L}-c_{R})^{\mathrm{loop}}= 12 m\, .
\end{equation}

\section{$U(1)_L$ charges of $\cN=2$ fields}\lab{ULapp}

In this appendix we discuss the $U(1)_L$ charges of the various fields in the $\cN=2$ quivers associated to a transverse ALE space in the A-series. Geometrically, this symmetry is realized as a $U(1)_L$ isometry of the ALE space corresponding to the A-series. This isometry is absent for the D- and E-series such that this $U(1)_L$ flavor symmetry is not present in the latter field theories. 

Recall that 4d $\cN=2$ quiver gauge theories we are interested in can be obtained by a quotient of $\cN=4$ super Yang--Mills. An $\cN=4$ vector multiplet consists of a gauge field $A^{\cN=4}_\mu$, six scalars $\varphi^i$ and fermions $\Psi^\cI_\alpha$, $\tilde \Psi_{\dot \alpha \cI}$, where $\cI=1, \dots, 4$ and $i=1, \dots, 6$.  They transform under $SO(1,3)_\ell \times SU(4)_R$ in the representations
\begin{equation}
A^{\cN=4}_\mu \in (\bs{2}, \bs 2, \bs 1)\, , \qquad \varphi^i \in (\bs 1, \bs 1, \bs 6)\, ,\qquad \Psi^\cI_\alpha \in (\bs 2, \bs 1, \bs 4)\, ,\qquad \tilde \Psi_{\dot \alpha \cI} \in (\bs 1, \bs 2, \bar{\bs 4})\, ,
\end{equation}
where $SU(4)_R$ is the R-symmetry of $\cN=4$ SYM. An $\cN=4$ vector multiplet can be decomposed into an $\cN=2$ vector multiplet and a hypermultiplet. Identifying $SU(2)_R \times U(1)_r\subset SU(4)_R$ as the $\cN=2$ R-symmetry, the $\cN=4$ vector multiplet decomposes  as
\begin{align}
SO(1,3)_\ell \times SU(4)_R \qquad &\to \qquad SO(1,3)_\ell \times  SU(2)_R \times U(1)_r\times U(1)_L\nn\\
A^{\cN=4}_\mu \in (\bs 2, \bs 2, \bs 1) \qquad&\to \qquad (\bs 2, \bs 2,\bs 1)_{0, 0}\nn\\
\varphi^i \in (\bs 1, \bs 1, \bs 6)\qquad & \to \qquad (\bs 1, \bs 1,  \bs 1)_{2,0}\oplus (\bs 1, \bs 1,  \bs 1)_{-2,0}\oplus (\bs 1, \bs 1, \bs 2)_{0,1}\oplus (\bs 1, \bs 1, \bs 2)_{0,-1}\nn\\
\Psi_{\alpha \cI}\in (\bs 2, \bs 1, \bs 4) \qquad & \to \qquad (\bs 2, \bs 1, \bs 1)_{1,1}\oplus (\bs 2, \bs 1, \bs 1)_{1,-1} \oplus(\bs 2, \bs 1, \bs 2)_{-1,0}\nn\\
\tilde \Psi^\cI_{\dot \alpha} \in (\bs 1, \bs 2, \bar{\bs 4}) \qquad &\to \qquad (\bs 1, \bs 2, \bs 1)_{-1,1}\oplus (\bs 1, \bs 2, \bs 1)_{-1,-1}\oplus(\bs 1, \bs 2, \bs 2)_{1,0}\, .\lab{leftdec}
\end{align}
The only states in the decomposition with non-vanishing $U(1)_L$ charge are the $(\bs 2, \bs 1, \bs 1)_{1,\pm,1}$ state in the decomposition of $\Psi_{\ax \cI}$, the $(\bs 1, \bs 2, \bs 1)_{-1, \pm 1}$ state from $\tilde \Psi_{\dot \ax}^\cI$, and finally the $(\bs 1, \bs 1, \bs 2)_{0, \pm 1}$ states from the decomposition of $\varphi^i$. Since the fermionic states $(\bs 2, \bs 1, \bs 1)_{1,\pm,1}$ and $(\bs 1, \bs 2, \bs 1)_{-1, \pm 1}$ are singlets under $SU(2)_R$, they are identified with the fermions in the $\cN=2$ hypermultiplet. Similarly, since the states $(\bs 1, \bs 1, \bs 2)_{0, \pm 1}$ are scalar doublets under $SU(2)_R$, we conclude that they comprise the scalar degrees of freedom in the $\cN=2$ hypermultiplet. In summary, the two scalars and the two fermions in an $\cN=2$ hypermultiplet have opposite charges $Q_L=\pm 1$ under $U(1)_L$. On the other hand the fields in the vector multiplet are uncharged under this $U(1)_L$ global symmetry. This then leads to the $U(1)_L$ charges shown in table \ref{2dspectrum}. Let us stress again that this additional global $U(1)_L$ symmetry only exists for the A-series. 

\newpage

\bibliographystyle{utcaps}
\bibliography{references}

\end{document}